\documentclass[%
 reprint,
 superscriptaddress,
 amsmath,amssymb,
 aps,
 pra,
 showkeys,
 nobalancelastpage,
]{revtex4-2}

\usepackage[utf8]{inputenc}
\usepackage[T1]{fontenc}

\usepackage{newtx}

\usepackage{bm}
\usepackage{xcolor}
\usepackage{graphicx}
\usepackage{dcolumn}
\usepackage{hyperref}

\DeclareMathOperator*{\argmin}{arg\,min}

\begin{document}

\title{Variational time reversal for free energy estimation in nonequilibrium steady states}

\author{Jorge L.\ Rosa-Ra\'ices}%
\email{jrosaraices@berkeley.edu}
\affiliation{%
  Department of Chemistry, University of California, Berkeley, California 94720, USA%
}
\author{David T.\ Limmer}%
\email{dlimmer@berkeley.edu}
\affiliation{%
  Department of Chemistry, University of California, Berkeley, California 94720, USA%
}
\affiliation{%
  Materials Science Division, Lawrence Berkeley National Laboratory, Berkeley, California 94720, USA%
}
\affiliation{%
  Chemical Science Division, Lawrence Berkeley National Laboratory, Berkeley, California 94720, USA%
}
\affiliation{%
  Kavli Energy NanoScience Institute, Berkeley, California 94720, USA%
} 

\begin{abstract}
 Studying the structure of systems in nonequilibrium steady states necessitates tools that quantify population shifts and associated deformations of equilibrium free energy landscapes under persistent currents.
 Within the framework of stochastic thermodynamics, we establish a variant of the Kawasaki--Crooks equality that relates nonequilibrium free energy corrections in overdamped Langevin systems to heat dissipation statistics along time-reversed relaxation trajectories computable with molecular simulation.
 Using stochastic control theory, we arrive at a general variational approach to evaluate the Kawasaki--Crooks equality, and use it to estimate distribution functions of order parameters in specific models of driven and active matter, attaining substantial improvement in accuracy over simple perturbative methods.
\end{abstract}


\maketitle

\section{Introduction}%
\label{sec:intro}

Free energy landscapes mold our modern understanding of structure throughout chemistry, materials science, and biology, by describing how energetics governs stability in the presence of thermal fluctuations~\cite{Tolman1938,Chandler1987}.
An assortment of computer-aided drug discovery~\cite{Muegge2023,Cournia2021,Mobley2012}, materials design~\cite{Lee2021,Kim2012,Rickman2002}, and basic science efforts are buttressed by well-worn free energy estimation methods~\cite{Chipot2023,Axelrod2022,York2023} that assume the systems targeted operate at or near thermodynamic equilibrium.
However, this assumption is conflicting with modern perspectives that seek to understand how nonequilibrium operating conditions govern the function of living, driven, and active systems.

The emerging nonequilibrium paradigm calls for new molecular simulation methods that can substantially extend the toolbox available at equilibrium to relevant classes of systems driven far from equilibrium.
Efforts to bridge this gap have lead to the development of stratification-based techniques for estimating likelihood surfaces along relevant order parameters, such as nonequilibrium umbrella sampling~\cite{Dickson2010}, forward flux sampling~\cite{Allen2009}, and weighted ensemble simulation~\cite{Zuckerman2017}.
Such techniques can produce likelihood estimates from flux balances on stratified simulation data, but choosing stratifications that expedite the convergence of likelihood estimates can be challenging in complex applications~\cite{hall2022practical}.
Moreover, these techniques operate independently and often distinctly from existing equilibrium methods. 
Especially in cases where one is interested in understanding how a free energy surface evolves as a system is driven progressively further from equilibrium, it would be advantageous to have a means to leverage existing equilibrium frameworks to aid in their evaluation. 

In this work, we develop nonequilibrium free energy surface estimators from results in stochastic thermodynamics, a rapidly developing theoretical framework that extends familiar thermodynamic notions away from equilibrium~\cite{Sekimoto2010,Roldan2014,Peliti2021,Shiraishi2024,Limmer2024}.
An important result in stochastic thermodynamics is the Kawasaki--Crooks equality~\cite[Chapter~5]{Crooks2001}, which we revisit in the current work from an optimal control viewpoint.
Our analysis yields a method for variationally exact free-energy estimation in nonequilibrium steady states, by evaluating changes in their statistics as they are driven away from a reference equilibrium system.
Our implementation draws from state-of-the-art approaches to compute dynamical large deviations~\cite{Chetrite2014,Das2019}, reaction rates~\cite{Das2022}, and local entropy flows in active systems~\cite{Boffi2024}.

The text is organized as follows.
Section~\ref{sec:TheoryAndMethod} develops the theory and outlines the computational approach of our proposed nonequilibrium FES estimation scheme. 
Section~\ref{sec:NumericalIllustrations} illustrates its application to simple models of driven and active matter.
Finally, Sec.~\ref{sec:Conclusions} provides concluding remarks.

\section{Theory and Method}%
\label{sec:TheoryAndMethod}

In this section, we employ notions from stochastic thermodynamics, path reweighing, and optimal control theory to arrive at estimators for how the free energy surface along collective coordinates deforms under nonequilibrium driving.
The estimators are amenable to evaluation via straightforward molecular simulation and incur modest post-processing overhead, as illustrated in Sec.~\ref{sec:NumericalIllustrations}.

\subsection{Nonequilibrium free energy surfaces}%
\label{ssec:TheoryAndMethod1}

Consider a $n$-dimensional system in thermodynamic equilibrium with a heat bath at temperature $\beta^{-1}$, its configuration $X = (x_1, \ldots, x_n)^\top$ distributed per the Boltzmann--Gibbs probability measure with density
\begin{equation}\label{eq:EquilibriumPDF}
  \rho^\mathrm{eq}(X) \,=\, \exp \, \bm[ \beta F^\mathrm{eq} - \beta U(X) \bm]
\end{equation}
induced by the potential energy $U$ on the configuration space $\Omega$.
The Helmholtz free energy
\begin{equation}\label{eq:EquilibriumFE} 
  F^\mathrm{eq}
    \,=\, \textstyle\int_\Omega \mathrm{d}X \, \rho^\mathrm{eq}(X) \, \bm[ U(X) + \beta^{-1} \ln \rho^\mathrm{eq}(X) \bm]
\end{equation}
is the minimum energy necessary to prepare the equilibrium system from a collection of independent degrees of freedom~\cite{gibbs1928collected}.
Simultaneously, $F^\mathrm{eq}$ is the global minimum of the free energy functional $\mathcal{F}_U$, which acts on an arbitrary nonequilibrium probability density $\rho^\mathrm{neq}$ supported on $\Omega$ to give
\begin{equation}\label{eq:FEFunctional}
\begin{aligned}
  \mathcal{F}_U(\rho^\mathrm{neq})
  \,&=\, \textstyle\int_\Omega \mathrm{d}X \, \rho^\mathrm{neq}(X) \, \bm[ U(X) + \beta^{-1} \ln \rho^\mathrm{neq}(X) \bm]
  \\ \,&=\, \mathcal{F}_U(\rho^\mathrm{eq}) + \beta^{-1} \mathcal{D}_{\mathrm{KL}}(\rho^\mathrm{neq} \,\Vert\, \rho^\mathrm{eq})
\end{aligned}
\end{equation}
where
\begin{equation}\label{eq:NonequilibriumFE}
  \mathcal{D}_{\mathrm{KL}}(\rho^\mathrm{neq} \,\Vert\, \rho^\mathrm{eq})
  \,\equiv\,
  \textstyle\int_\Omega \mathrm{d}X \, \rho^\mathrm{neq}(X) \ln \dfrac{\rho^\mathrm{neq}(X)}{\rho^\mathrm{eq}(X)}
  \,\ge\, 0
\end{equation}
is the Kullback--Leibler divergence from equilibrium of the state specified by $\rho^\mathrm{neq}$.
If the latter is a physically realizable nonequilibrium state of the system with potential energy $U$, then $\beta^{-1}\mathcal{D}_\mathrm{KL}(\rho^\mathrm{neq} \,\Vert\, \rho^\mathrm{eq})$ is the nonequilibrium free energy extractable from the system as it autonomously regresses to thermal equilibrium with the enclosing heat bath~\cite{Gaveau2002,Sivak2012}.

Although insightful as a scalar measure of deviation from equilibrium, the nonequilibrium free energy in Eq.~\eqref{eq:NonequilibriumFE} informs little about the structure and conformational stability of a driven system.
However, the free-energy \emph{surface} (FES) has long been useful to assess high-dimensional phenomena in terms of low-dimensional structural features that capture the governing physics~\cite{kirkwood1935statistical,Wales2006}.
While away from equilibrium such a surface does not encode the same response properties as in equilibrium, it still relays the stability of conformations, or groups of configurations, relative to one another along prescribed order parameters~\cite{baiesi2013update}.
For an order parameter $R : \Omega \to \mathbb{R}$ mapping onto scalar conformations of a system with configuration density $\rho(X)$, equilibrium or otherwise, the FES 
at conformation $r$ is simply
\begin{equation}\label{eq:FESurface}
  \beta F(r) \,\equiv\, -\ln P(r) \,=\, -\ln \textstyle\int_{\Omega} \mathrm{d}X \, \delta \bm( R(X) - r \bm) \, \rho(X)
\end{equation}
where $P(r)$ is the likelihood of observing the system in conformation $r$.
Numerical methods to accurately estimate the equilibrium FES 
$F^\mathrm{eq}(r)$ up to an additive constant are widespread and readily applicable to complex systems~\cite{Torrie1977,Laio2008,Comer2015,Lelievre2010,frenkel2023understanding}. The extension to nonequilibrium applications entails identifying effective estimators of the FES 
shift
\begin{equation}\label{eq:FESurfaceShift}
  \beta F^\mathrm{neq}(r) - \beta F^\mathrm{eq}(r) \,=\, - \ln \dfrac{P^\mathrm{neq}(r)}{P^\mathrm{eq}(r)}
\end{equation}
where $P^\mathrm{neq}(r)$ is the likelihood of conformation $r$ for a system in its nonequilibrium steady state (NESS), and $P^\mathrm{eq}(r)$ is the likelihood of the same conformation absent a driving force. Methods to estimate this shift efficiently are lacking.

\subsection{Kawasaki--Crooks equality for $\beta F^\mathrm{neq}(r)$}%
\label{ssec:TheoryAndMethod2}

The nonequilibrium FES 
shift in Eq.~\eqref{eq:FESurfaceShift} can be challenging to estimate for complex systems.
Whereas $P^\mathrm{eq}(r)$ can be efficiently evaluated via statistical reweighing of equilibrium configuration data that is importance-sampled around $r$, a similar strategy for evaluating $P^\mathrm{neq}(r)$ is all but intractable absent a closed-form nonequilibrium analog of Eq.~\eqref{eq:EquilibriumPDF}.
Working within the framework of stochastic thermodynamics, in this section we derive an alternative expression for the nonequilibrium free energy surface shift that transfers the burden of nonequilibrium importance sampling away from configuration space and onto path space, where path sampling methods can be leveraged to great effect as we show in later sections.

We develop our results assuming an overdamped diffusive dynamics with additive noise, but considering to the generality of the theoretical foundation of our arguments~\cite{Pra1991,Crooks2001,Chetrite2008,Gawedzki2013}, we surmise our results can be extended to other classical Markovian dynamics in common use throughout the stochastic thermodynamics literature.
In our setting, the system's equilibrium measure with configuration density $\rho^\mathrm{eq}$ is preserved by the stochastic differential equation (SDE) with unit mobility
\begin{equation}\label{eq:EquilibriumSDE}
  \mathrm{d}X_t =\, -\nabla U (X_t) \, \mathrm{d}t +\! \sqrt{2\smash[t]{\beta^{-1}}} \mathrm{d} W_t
\end{equation}
where the standard $n$-dimensional Wiener process $(W_t)_{t \ge 0}$ encodes the system's coupling to its enclosing heat bath 
and satisfies $\langle W_t \rangle = 0$ and $\langle W_t \cdot W_s \rangle = n\min(t, s)$ for $t, s \ge 0$.
Likewise, the nonequilibrium steady state density $\rho^\mathrm{neq}$ corresponds to the invariant measure of the SDE
\begin{equation}\label{eq:NonequilibriumSDE}
  \mathrm{d}X_t =\, \bm[ f(X_t) - \nabla U (X_t) \bm] \, \mathrm{d}t +\! \sqrt{2\smash[t]{\beta^{-1}}} \mathrm{d} W_t
\end{equation}
where the vector field $f$, which encodes the externally or autonomously generated forces that drive the system out of equilibrium, is assumed to satisfy $\nabla \cdot f = 0$ throughout $\Omega$, ensuring that $\rho^\mathrm{neq}$ is not a Boltzmann--Gibbs density~\cite{Risken1996}.
Unless otherwise specified, the SDEs in Eqs.~\eqref{eq:EquilibriumSDE} and~\eqref{eq:NonequilibriumSDE} are both endowed with identical equilibrium initial conditions $X_0 \sim \rho^\mathrm{eq}$ throughout our discussion.

At any time horizon $T$, the SDEs in Eqs.~\eqref{eq:EquilibriumSDE} and~\eqref{eq:NonequilibriumSDE} respectively induce mutually overlapping probability measures $\mathbb{P}^\mathrm{eq}$ and $\mathbb{P}^\mathrm{neq}$ on path space with 
relative density~\cite{Maes2008}
\begin{equation}\label{eq:RelativeDensity}
\begin{aligned}
  \dfrac{\mathrm{d}\mathbb{P}^\mathrm{neq}(\boldsymbol{X})}{\mathrm{d}\mathbb{P}^\mathrm{eq}(\boldsymbol{X})}
  \,&=\, \exp \, \bm[ -\beta \mathcal{S}_T (\boldsymbol{X}) \bm]
  \\
  \,&=\, \exp \, \bm\{ -\beta \, \bm[ \mathcal{T}_T(\boldsymbol{X}) + \mathcal{Q}_T(\boldsymbol{X}) \bm] / 2 \bm\}
\end{aligned}
\end{equation}
at every sample-path $\boldsymbol{X} = (X_t)_{0\le t\le T}$.
The path functionals
\begin{subequations}\label{eq:RelativeDensityTerms}
\begin{align}
  \mathcal{T}_T(\boldsymbol{X}) \,&=\, \textstyle\int_0^T \mathrm{d}t \, \tfrac{1}{2} | f(X_t) |^2 - \nabla U(X_t) \cdot f(X_t)%
  \label{eq:Traffic} \\
  \mathcal{Q}_T(\boldsymbol{X}) \,&=\, -\textstyle\int_0^T \mathrm{d}X_t \circ f(X_t)%
  \label{eq:Heat}
\end{align}
\end{subequations}
are called the traffic and the excess heat~\cite{baiesi2009fluctuations}, respectively, with the stochastic integral in the latter functional interpreted in the Stratonovich sense~\cite[Chapter~3.2]{Pavliotis2014}. This decomposition separates the change of path measure into symmetric and antisymmetric components under the pathwise time-involution operation $\boldsymbol{X} \mapsto \tilde{\boldsymbol{X}} \!= (X_{T-t})_{0\le t\le T}$, as in
\begin{equation}\label{eq:RelativeDensityTermsTimeInvolution}
  \mathcal{T}_T(\tilde{\boldsymbol{X}}) \,=\, \mathcal{T}_T(\boldsymbol{X})
  \;\;\text{and}\;\;
  \mathcal{Q}_T(\tilde{\boldsymbol{X}}) \,=\, -\mathcal{Q}_T(\boldsymbol{X}) \,,
\end{equation}
clarifying the role of kinetic quantities in path reweighing theories~\cite{gao2019nonlinear,limmer2021large}.
The relative density in Eq.~\eqref{eq:RelativeDensity} between equilibrium and nonequilibrium path measures thus exhibits a time-involution symmetry through the stochastic action $\mathcal{S}_T(\boldsymbol{X})$, namely
\begin{equation}\label{eq:RelativeDensityTimeInvolution}
  \exp \, \bm[ -\beta \mathcal{S}_T (\tilde{\boldsymbol{X}}) \bm] \,=\, \exp \bm \{ -\beta \bm[ \mathcal{S}_T(\boldsymbol{X}) - \mathcal{Q}_T(\boldsymbol{X}) \bm] \bm \} \,,
\end{equation}
which is an instance of a detailed fluctuation theorem~\cite{Crooks1999,Seifert2012}.

The above facts allow us to derive a Kawasaki--Crooks equality relating the FES 
shift $\beta F^\mathrm{neq}(r) - \beta F^\mathrm{eq}(r)$ in Eq.~\eqref{eq:FESurfaceShift} to an average over nonequilibrium paths starting at conformation $r$\,~\cite{yamada1967nonlinear,morriss1985isothermal,crooks2000path}.
The derivation detailed in Appendix~\ref{sec:AppendixA} yields
\begin{equation}\label{eq:KawasakiCrooks}
  \ln \dfrac{P^\mathrm{neq}(r)}{P^\mathrm{eq}(r)}
  \,=\,
  \lim_{T \to \infty} \ln \dfrac%
  {\langle \exp \, \bm[ \beta \mathcal{Q}_T(\boldsymbol{X}) \bm] \rangle^\mathrm{neq}_r}%
  {\langle \exp \, \bm[ \beta \mathcal{Q}_T(\boldsymbol{X}) \bm] \rangle^\mathrm{neq}}
\end{equation}
where the path average $\langle \cdot \rangle_r^\mathrm{neq}$ is evaluated against realizations of the nonequilibrium SDE in Eq.~\eqref{eq:NonequilibriumSDE} initialized per the constrained equilibrium density
\begin{equation}\label{eq:RestrictedEquilibriumPDF}
  \rho_r^\mathrm{eq}(X) \,=\, \delta \bm( R(X) - r \bm) \, \rho^\mathrm{eq}(X) / P^\mathrm{eq}(r)
\end{equation}
The path average in the denominator of Eq.~\eqref{eq:KawasakiCrooks} satisfies the integral fluctuation relation
\begin{equation}\label{eq:IFT}
  1 \,=\, \langle \exp \, \bm[ \beta \mathcal{Q}_T(\boldsymbol{X}) \bm] \rangle^\mathrm{neq} \;\;\text{for all}\;\; T > 0
\end{equation}
for the excess heat $\mathcal{Q}_T$, also derived in Appendix~\ref{sec:AppendixA}.
Comparing Eq.~\eqref{eq:KawasakiCrooks} with Eq.~\eqref{eq:FESurfaceShift}, we observe that the Kawasaki--Crooks equality effectively replaces the 
constrained nonequilibrium average in the definitional FES 
shift with a path-integral reweighing of its equilibrium counterpart.

Though more computationally appealing than Eq.~\eqref{eq:FESurfaceShift}, the Kawasaki--Crooks FES 
shift in Eq.~\eqref{eq:KawasakiCrooks} involves exponential averages of path observables with time-extensive fluctuations, which can be challenging to estimate despite substantial advancements in enhanced path-sampling methods~\cite{Gingrich2015,Nemoto2016,Ferre2018,Ferre2021}.
Through cumulant expansions and path reweighing, we may unpack Eq.~\eqref{eq:KawasakiCrooks} in ways that reveal physical insight and reduce numerical variance.
For instance, the right-hand side may be expanded in cumulants of the excess heat $\beta\mathcal{Q}_T$ to obtain, at leading order,
\begin{equation}\label{eq:KawasakiCrooksPerturbativeNonequilibrium}
\begin{aligned}
  \ln \dfrac{P^\mathrm{neq}(r)}{P^\mathrm{eq}(r)}
  \,&\approx\,
  \lim_{T \to \infty} \bm[%
  \langle \beta \mathcal{Q}_T(\boldsymbol{X}) \rangle^\mathrm{neq}_r -%
  \langle \beta \mathcal{Q}_T(\boldsymbol{X}) \rangle^\mathrm{neq}%
  \bm]
  \\[2pt]
  \,&\equiv\, \beta F^\mathrm{eq}(r) - \beta \hat{F}^\mathrm{neq}_{\mathcal{Q}}(r)
\end{aligned}
\end{equation}
where $\hat{F}^\mathrm{neq}_{\mathcal{Q}}(r)$ is a heat-based estimate of the nonequilibrium FES. 
To leading order, then, nonequilibrium free energy shifts arise at conformations $r$ from which a substantial amount of excess heat is dissipated as the system is driven from equilibrium towards its nonequilibrium steady state~\cite{morriss1985isothermal,Sivak2012,kuznets2021dissipation,kuznets2023inferring}.
Alternatively, reweighing the averages in Eq.~\eqref{eq:KawasakiCrooks} into the equilibrium path ensemble with measure $\mathbb{P}^\mathrm{eq}$, and subsequently expanding up to leading order in the traffic $\beta\mathcal{T}_T$, obtains
\begin{equation}\label{eq:KawasakiCrooksPerturbativeEquilibrium}
\begin{aligned}
  \ln \dfrac{P^\mathrm{neq}(r)}{P^\mathrm{eq}(r)}
  \,&=\,
  \lim_{T \to \infty} \ln \dfrac%
  {\langle \exp \, \bm[ -\beta (\mathcal{S}_T - \mathcal{Q}_T)(\boldsymbol{X}) \bm] \rangle^\mathrm{eq}_r}%
  {\langle \exp \, \bm[ -\beta (\mathcal{S}_T - \mathcal{Q}_T)(\boldsymbol{X}) \bm] \rangle^\mathrm{eq}}
  \\[2pt]
  \,&\approx\,
  \lim_{T \to \infty} \bm[%
  \langle \beta\mathcal{T}_T(\boldsymbol{X}) \rangle^\mathrm{eq} -%
  \langle \beta\mathcal{T}_T(\boldsymbol{X}) \rangle^\mathrm{eq}_r%
  \bm] / 2
  \\[2pt]
  \,&\equiv\, \beta F^\mathrm{eq}(r) - \beta \hat{F}^\mathrm{neq}_{\mathcal{T}}(r)
\end{aligned}
\end{equation}
where $\hat{F}^\mathrm{neq}_{\mathcal{T}}(r)$ is traffic-based estimate of the nonequilibrium FES. 
This result is related to nonlinear response theory for NESSs~\cite{gao2019nonlinear,lesnicki2020field,limmer2021large} and has been used to estimate nonequilibrium population shifts with equilibrium averages of time-reversal symmetric trajectory observables~\cite{lesnicki2021molecular}.

\subsection{Variational time reversal estimation of $\beta F^\mathrm{neq}(r)$}%
\label{ssec:TheoryAndMethod3}

FES estimates $\hat{F}^\mathrm{neq}_{\mathcal{Q}}(r)$ in Eq.~\eqref{eq:KawasakiCrooksPerturbativeNonequilibrium} and $\hat{F}^\mathrm{neq}_{\mathcal{T}}(r)$ in Eq.~\eqref{eq:KawasakiCrooksPerturbativeEquilibrium}, although appealing for their simplicity, can yield highly biased estimates of the Kawasaki--Crooks FES shift in Eq.~\eqref{eq:KawasakiCrooks} under strong driving~\cite{Hummer2007}.
Stochastic optimal control arguments yield a variational estimator for the exponential average in Eq.~\eqref{eq:KawasakiCrooks} by introducing a guiding field or control policy $\varv_{\boldsymbol{\theta}}$ into the nonequilibrium dynamics, resulting in
\begin{equation}\label{eq:KawasakiCrooksVariational}
\begin{aligned}
  & \lim_{T \to \infty} \ln \,
  \langle \exp \, \bm[\beta \mathcal{Q}_T(\boldsymbol{X}) \bm] \rangle_r^\mathrm{neq}
  \\[-4pt]
  & \qquad\,=\,
  \lim_{T \to \infty} \ln \,
  \Bigl\langle%
    \exp \, \bm[ \beta \mathcal{Q}_T(\boldsymbol{X}) \bm]%
    \dfrac{\mathrm{d}\mathbb{P}^\mathrm{neq}(\boldsymbol{X})}{\mathrm{d}\mathbb{P}^{\varv_{\boldsymbol{\theta}}}(\boldsymbol{X})}%
  \Bigr\rangle_r^{\varv_{\boldsymbol{\theta}}}
  \\
  & \qquad\,=\,
  \lim_{T \to \infty} \sup_{\{\varv_{\boldsymbol{\theta}}\}} \, \beta%
  \Bigl\langle%
    \mathcal{Q}_T(\boldsymbol{X}) - \textstyle\int_0^T \!\mathrm{d}t \, |\varv_{\boldsymbol{\theta}}(X_t)|^2%
  \Bigl\rangle_r^{\varv_{\boldsymbol{\theta}}}
\end{aligned}
\end{equation}
The first equality introduces a new path measure $\mathbb{P}^{\varv_{\boldsymbol{\theta}}}$ and associated expectation $\langle \cdot \rangle_r^{\varv_{\boldsymbol{\theta}}}$ governed by the controlled SDE
\begin{equation}\label{eq:ControlledNonequilibriumSDE}
  \mathrm{d}X_t =\, \bm[ f(X_t) - \nabla U(X_t) - 2\varv_{\boldsymbol{\theta}}(X_t) \bm] \, \mathrm{d}t +\! \sqrt{2\smash[t]{\beta^{-1}}} \mathrm{d}W_t
\end{equation}
endowed with the initial condition $X_0 \sim \rho_r^\mathrm{eq}$.
The change of path measure from $\mathbb{P}^\mathrm{neq}$ to $\mathbb{P}^{\varv_{\boldsymbol{\theta}}}$ incurs the 
relative density
\begin{equation}\label{eq:ControlledRelativeDensity}
\begin{aligned}
  \dfrac{\mathrm{d}\mathbb{P}^\mathrm{neq}(\boldsymbol{X})}{\mathrm{d}\mathbb{P}^{\varv_{\boldsymbol{\theta}}}(\boldsymbol{X})}
  \,=\,
  \exp \,
  &\bm{\Bigl[}%
    \,-\beta \textstyle\int_0^T \mathrm{d}t \, |\varv_{\boldsymbol{\theta}}(X_t)|^2%
  \\%
  &\phantom{\bm{\Bigl[}}+%
    \sqrt{2 \beta} \textstyle\int_0^T \mathrm{d}W_t \cdot \varv_{\boldsymbol{\theta}}(X_t)%
  \bm{\Bigr]}
\end{aligned}
\end{equation}
at every sample-path $\boldsymbol{X}$ of Eq.~\eqref{eq:ControlledNonequilibriumSDE}~\cite[Theorem~8.6.6]{Oksendal2003}.
The second equality states an optimal control result, asserting the existence of a lower bound on the exponential average as an supremum over controlled nonequilibrium trajectory ensembles indexed by the control policy $\varv_{\boldsymbol{\theta}}$, of which a unique choice saturates the supremum in Eq.~\eqref{eq:KawasakiCrooksVariational}~\cite{Bierkens2014}.
The variationally optimal estimator affords statistically unbiased estimation of the nonequilibrium FES 
shift and additionally minimizes statistical variance across controlled estimators~\cite{Thijssen2015}.

\begin{figure*}[t]
\centering
  \includegraphics[width=\textwidth]%
    {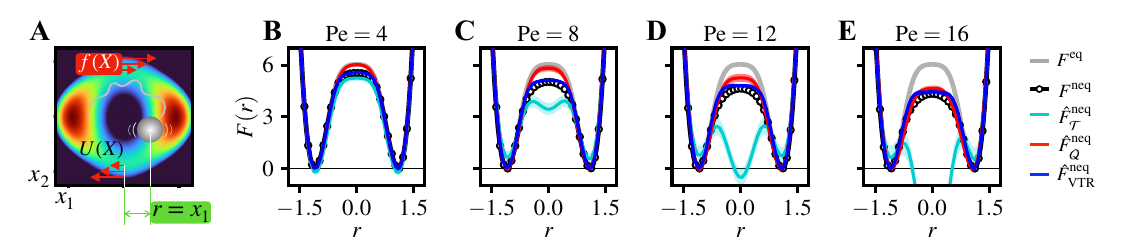}
\caption{%
  \textbf{A}.
  Two-dimensional Brownian particle in the potential $U(X)$ [Eq.~\eqref{eq:ShearedParticlePES}] undergoes linear shear $f(X)$ [Eq.~\eqref{eq:ShearedParticleNonequilibriumDrift}] at rate $\mathrm{Pe}$.
  \textbf{B}--\textbf{E}.
  Nonequilibrium free energy surfaces along order parameter $r = x_1$ (identified in panel~A) at shear rates $\mathrm{Pe} = 4, 8, 12, 16$ evaluated with various estimators (colored lines) are shown together with the reference equilibrium free energy surface $F^\mathrm{eq}(r)$ (gray line).
  The black circles evaluate the numerically exact nonequilibrium free energy surface $F^\mathrm{neq}(r)$ from histogram data at each shear rate.
}\label{fig:sheared_particle_schematic_and_feps}
\end{figure*}

In practice, online reinforcement learning methods~\cite{Sutton2018,rose2021reinforcement,Das2021b,yan2022learning} allow optimizing $\varv_{\boldsymbol{\theta}}$ to approach the supremum in Eq.~\eqref{eq:KawasakiCrooksVariational} by gradient ascent of the reward function
\begin{equation}\label{eq:KawasakiCrooksVariationalReward}
  \mathsf{R}(\boldsymbol{\theta}) \,=\,
    \lim_{T \to \infty} \bm[%
      \langle \mathcal{R}_T(\boldsymbol{X}, \boldsymbol{\theta}) \rangle^{\varv_{\boldsymbol{\theta}}}_r -%
      \langle \mathcal{R}_T(\boldsymbol{X}, \boldsymbol{\theta}) \rangle^{\varv_{\boldsymbol{\theta}}}%
    \bm]
\end{equation}
with respect to policy parameters $\boldsymbol{\theta}$, where
\begin{equation}\label{eq:KawasakiCrooksVariationalPathwiseReward}
  \mathcal{R}_T(\boldsymbol{X}, \boldsymbol{\theta}) \,=\,
    \beta \mathcal{Q}_T(\boldsymbol{X}) + \dfrac{\mathrm{d}\mathbb{P}^\mathrm{neq}(\boldsymbol{X})}{\mathrm{d}\mathbb{P}^{\varv_{\boldsymbol{\theta}}}(\boldsymbol{X})}
\end{equation}
is the pathwise reward.
However, stochastic estimation of the reward gradient
\begin{equation}\label{eq:KawasakiCrooksVariationalRewardGradient}
\begin{aligned}
  \nabla_{\boldsymbol{\theta}} \mathsf{R}(\boldsymbol{\theta}) \,=\,
    \lim_{T \to \infty} \langle \bm{\mathcal{M}}_T(\boldsymbol{X}, \boldsymbol{\theta})\, \bm[ &\mathcal{R}_T(\boldsymbol{X}, \boldsymbol{\theta}) \\ &- \langle \mathcal{R}_T(\boldsymbol{X}, \boldsymbol{\theta}) \rangle^{\varv_{\boldsymbol{\theta}}} \bm] \rangle_r^{\varv_{\boldsymbol{\theta}}}
\end{aligned}
\end{equation}
where
\begin{equation}\label{eq:MalliavinWeight}
  \bm{\mathcal{M}}_T(\boldsymbol{X}, \boldsymbol{\theta}) \,=\,
    \sqrt{2\beta} \textstyle\int_0^T \!\mathrm{d}W_t \cdot \nabla_{\boldsymbol{\theta}} \varv_{\boldsymbol{\theta}}(X_t)
\end{equation}
is the pathwise Malliavin weight~\cite{Warren2014}, can be daunting in the relevant long-time limit due to a confluence of time-extensive fluctuations in
$\mathcal{R}_T$ and in $\bm{\mathcal{M}}_T$.

Due to these considerations, we eschew online reinforcement learning of the optimal policy that saturates the supremum in Eq.~\eqref{eq:KawasakiCrooksVariational} for general order parameters, in favor of a suboptimal but effective policy that can be estimated through offline regression on readily obtained NESS trajectory data; a similar approach has been used in related rare-event problems~\cite{singh2023variational,jung2023machine}.
In particular, we target the control policy that saturates the supremum in Eq.~\eqref{eq:KawasakiCrooksVariational} for the \emph{identity} order parameter such that $R(X) = X$, which takes the form
\begin{equation}\label{eq:FokkerPlanckVelocity}
  \varv^\mathrm{neq}(X) \,=\, f(X) - \beta^{-1} \nabla \ln \, \bm[ \rho^\mathrm{neq}(X) / \rho^\mathrm{eq}(X) \bm]
\end{equation}
and is derived in Appendix~\ref{sec:AppendixB}.
We recognize $\varv^\mathrm{neq}$ as the velocity field associated with the NESS configuration density $\rho^\mathrm{neq}$ through the steady-state Fokker--Planck equation
\begin{equation}\label{eq:FokkerPlanckEquation}
  0 \,=\, \bm(\mathscr{L}^\dag \rho^\mathrm{neq} \bm)(X)
    \,=\, -\nabla \cdot \bm( \rho^\mathrm{neq} \, \varv^\mathrm{neq} \bm)(X)
    \; \forall \; X \in \Omega
\end{equation}
with $\mathscr{L}^\dag$ the $\mathrm{L}^2(\Omega)$-adjoint of the generator corresponding to the SDE in Eq.~\eqref{eq:NonequilibriumSDE}~\cite[Chapter~4.1]{Pavliotis2014}.

Control policy $\varv^\mathrm{neq}$ is generally suboptimal for order parameters that encode surjective (i.e., many-to-one) configuration mappings, as it fails to attain the supremum in Eq.~\eqref{eq:KawasakiCrooksVariational} as follows from applying Jensen's inequality~\cite[Theorem~1.6.2]{Durrett2010}.
Indeed,
\begin{equation}\label{eq:KawasakiCrooksVariationalJensen}
\begin{aligned}
  \lim_{T \to \infty} & \ln%
  \langle \exp \, \bm[ \beta \mathcal{Q}_T(\boldsymbol{X}) \bm] \rangle_r^\mathrm{neq}
  \\ &\quad\ge\,
  {\textstyle\int_\Omega \mathrm{d}X \, \rho_r^\mathrm{eq}(X)} \lim_{T \to \infty}
  \ln%
  \langle \exp \, \bm[ \beta \mathcal{Q}_T(\boldsymbol{X}) \bm] \rangle_X^\mathrm{neq}
  \\[4pt] &\quad=\,
  {\textstyle\int_\Omega \mathrm{d}X \, \rho_r^\mathrm{eq}(X)} \lim_{T \to \infty} \sup_{\{\varv_{\boldsymbol{\theta}}\}} \, 
  \langle \mathcal{R}_T(\boldsymbol{X}, \boldsymbol{\theta}) \rangle_X^{\varv_{\boldsymbol{\theta}}}
  \\ &\quad\equiv\, \beta F^\mathrm{eq}(r) - \beta \hat{F}^\mathrm{neq}_\mathrm{VTR}(r)
\end{aligned}
\end{equation}
with the inequality saturated only for invertible (i.e., one-to-one \emph{and} onto) order parameters, and with the supremum attained at all $X \in \Omega$ up to an additive constant---stated in Appendix~\ref{sec:AppendixB}---at the control policy $\varv_{\boldsymbol{\theta}} = \varv^\mathrm{neq}$ in Eq.~\eqref{eq:FokkerPlanckVelocity}.
Despite this shortcoming, the central advantage of control policy $\varv^\mathrm{neq}$ is that it can be estimated without recourse to reinforcement learning methods, for instance by fitting NESS trajectory data through a steady-state velocity matching problem
\begin{equation}\label{eq:FokkerPlanckVelocityMSE}
  \varv^\mathrm{neq}
    \,=\,
    \argmin_{\{\varv_{\boldsymbol{\theta}}\}} \lim_{T \to \infty} \tfrac{1}{T} \Bigl\langle \textstyle\int_0^T \!\mathrm{d}t \, |\varv^\mathrm{neq}(X_t) - \varv_{\boldsymbol{\theta}}(X_t)|^2 \Bigr\rangle^\mathrm{neq}
\end{equation}
where $\varv_{\boldsymbol{\theta}}$ is a suitable control policy ansatz.

In what follows, we work with ans\"{a}tze $\varv_{\boldsymbol{\theta}}$ of the form
\begin{equation}\label{eq:FokkerPlanckVelocityAnsatz}
  \varv_{\boldsymbol{\theta}}(X) \,=\, f(X) - \nabla V_{\boldsymbol{\theta}}(X)
\end{equation}
where the control potential $V_{\boldsymbol{\theta}}$ is represented with a differentiable ansatz.
While a deep-learning solution to the steady-state velocity matching problem in Eq.~\eqref{eq:FokkerPlanckVelocityMSE} can most accurately estimate $\varv^\mathrm{neq}$~\cite{Boffi2023,Boffi2024}, our numerical tests show that simple, physically transparent parametrizations of the control potential $V_{\boldsymbol{\theta}}$ can yield substantial bias reduction in a variational estimate of the FES shift using the controlled estimator $\hat{F}^\mathrm{neq}_\mathrm{VTR}(r)$ introduced in Eq.~\eqref{eq:KawasakiCrooksVariationalJensen}, relative to perturbative estimates using $\hat{F}^\mathrm{neq}_{\mathcal{Q}}(r)$ in Eq.~\eqref{eq:KawasakiCrooksPerturbativeNonequilibrium} and $\hat{F}^\mathrm{neq}_{\mathcal{T}}(r)$ in Eq.~\eqref{eq:KawasakiCrooksPerturbativeEquilibrium}, at the minimal computational overhead of solving a linear system of equations as detailed in Appendix~\ref{sec:AppendixC}.

\begin{figure*}[t]
\centering
\includegraphics[width=\textwidth]%
  {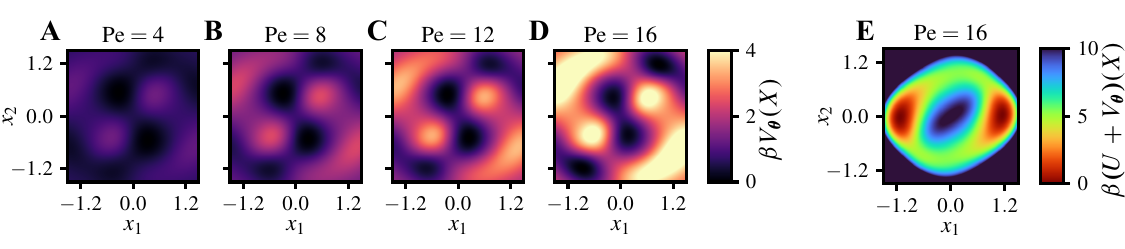}
\caption{%
  \textbf{A}--\textbf{D}.
  Optimized control potentials $\beta V_{\boldsymbol{\theta}}(X)$ [Eq.~\eqref{eq:ShearedParticleControlPotentialAnsatz}] associated with the control policies used to evaluate $\hat{F}^\mathrm{neq}_\mathrm{VTR}(r)$ in Figs.~\ref{fig:sheared_particle_schematic_and_feps}B-E at the respective shear rates $\mathrm{Pe}$.
  \textbf{E}.
  The optimized surface $\beta \bm( U(X) + V_{\boldsymbol{\theta}}(X) \bm)$ at $\mathrm{Pe} = 16$, with $U(X)$ from Eq.~\eqref{eq:ShearedParticlePES}, corresponds to the potential energy of a system whose equilibrium distribution encodes the nonequilibrium statistics of the sheared Brownian particle.
}\label{fig:sheared_particle_control_policy}
\end{figure*}

\section{Numerical illustrations}%
\label{sec:NumericalIllustrations}

In this section, we discuss implementation details and illustrate the accuracy of estimators of $\beta F^\mathrm{neq}(r)$ introduced in the previous section, including the perturbative estimators $\beta \hat{F}^\mathrm{neq}_{\mathcal{Q}}(r)$ and $\beta \hat{F}^\mathrm{neq}_{\mathcal{T}}(r)$, and the variational time reversal estimator $\beta \hat{F}^\mathrm{neq}_\mathrm{VTR}(r)$.
We consider a low-dimensional model where numerically exact results can be obtained, and a high-dimensional active filament where our methodology can lend insight into an activity driven collapse.
We set $\beta = 1$ and work with dimensionless units in both numerical examples.

\subsection{Sheared particle in confined potential}%
\label{ssec:ShearedParticle}

As a simple model to compare nonequilibrium FES 
estimators, we consider the two-dimensional Brownian particle governed by nonequilibrium SDE in Eq.~\eqref{eq:NonequilibriumSDE} with potential
\begin{equation}\label{eq:ShearedParticlePES}
  U(X) \,=\, 12 + 8 x_1^4 - 12 x_2^2 + 6 x_2^4 + 20 x_1^2 (x_2^2 - 1)
\end{equation}
and linear shear drift
\begin{equation}\label{eq:ShearedParticleNonequilibriumDrift}
  f(X) \,=\, \mathrm{Pe} \; (x_2 ,\, 0)^\top
\end{equation}
where $\mathrm{Pe}$ is the shear rate. This system is illustrated in Fig.~\ref{fig:sheared_particle_schematic_and_feps}A.
At each $\mathrm{Pe} \!\neq\! 0$, the particle evolves toward a nonequilibrium steady state with analytically unknown probability density $\rho^\mathrm{neq}(X)$.
We characterize the associated nonequilibrium FES 
along the $x_1$ coordinate through the estimators of the Kawasaki--Crooks relation in Eq.~\eqref{eq:KawasakiCrooks} that were introduced in Section~\ref{sec:TheoryAndMethod}.

FES estimates were evaluated on a uniform grid along the order parameter coordinate, composed of $48$ equidistant points in the interval $[-1.95, 1.95]$.
At each grid point $r$, an ensemble of approximately $10^4$ particle configurations conditioned on $x_1\!= r$ was generated using an inverse transform sampler of the restricted equilibrium distribution with the density $\rho^\mathrm{eq}_r$ in Eq.~\eqref{eq:RestrictedEquilibriumPDF}.
Trapezoidal integration was used to estimate the equilibrium FES $F^\mathrm{eq}(r)$ directly from a histogram of the equilibrium probability density function.
To estimate the nonequilibrium FES $F^\mathrm{neq}(r)$ at shear rates $\mathrm{Pe} \!\in\! \{4, 8, 12, 16\}$ using the perturbative estimators in Eqs.~\eqref{eq:KawasakiCrooksPerturbativeEquilibrium}, \eqref{eq:KawasakiCrooksPerturbativeNonequilibrium}, and~\eqref{eq:KawasakiCrooksVariationalJensen}, at least $10^4$ configurations sampled from the conditional equilibrium ensemble at $r$ were evolved through numerical integration of the SDEs in Eqs.~\eqref{eq:EquilibriumSDE}, \eqref{eq:NonequilibriumSDE}, and \eqref{eq:ControlledNonequilibriumSDE}, respectively, over a long enough time interval to be distributed per the corresponding steady state.
Whereas trajectories up to $80$ time units long were required to relax into the nonequilibrium steady states of the SDEs in Eqs.~\eqref{eq:NonequilibriumSDE} and~\eqref{eq:ControlledNonequilibriumSDE} across all order parameters values and shear rates, at least $300$ time units were needed for relaxation from each conditioned equilibrium ensemble 
into the unconditioned equilibrium ensemble 
through the equilibrium SDE in Eq.~\eqref{eq:EquilibriumSDE}.
The Euler--Maruyama method~\cite{Kloeden1992} was used to integrate all SDEs with a stepsize of $5 \times 10^{-3}$ time units.

Deploying the estimators defined in Eqs.~\eqref{eq:KawasakiCrooksPerturbativeNonequilibrium}, \eqref{eq:KawasakiCrooksPerturbativeEquilibrium}, and~\eqref{eq:KawasakiCrooksVariationalJensen} respectively yields the red, cyan, and blue free energy surfaces in Figs.~\ref{fig:sheared_particle_schematic_and_feps}B-E.
Also plotted is the empirically estimated nonequilibrium free energy surface, which is our ground truth for qualitative comparison with the perturbative and variational estimates.
We find that the effect of an increasing shear rate $\mathrm{Pe}$ is to suppress the barrier height relative to the equilibrium free energy surface, while leaving the structure around the metastable points largely invariant.
The traffic-based free energy estimator $\hat{F}^\mathrm{neq}_{\mathcal{T}}(r)$ is accurate at low $\mathrm{Pe}$ but rapidly deteriorates as $\mathrm{Pe}$ is increased.
For $\mathrm{Pe}$ greater than $4$, this estimator erroneously predicts a metastable state at the top of the barrier, whose stability grows with increasing $\mathrm{Pe}$.
Comparatively, the heat-based estimator $\hat{F}^\mathrm{neq}_{\mathcal{Q}}(r)$ shows better agreement with the ground truth barrier height across the range of shear rates considered, as well as the qualitative form of $\hat{F}^\mathrm{neq}(r)$.
It is fortuitously accurate at the largest $\mathrm{Pe}$ considered, though less accurate at intermediate values.

The most accurate estimator with regards to barrier height is the variational estimator $\hat{F}^\mathrm{neq}_\mathrm{VTR}(r)$, though the estimate qualitatively deviates from the ground truth barrier shape at higher $\mathrm{Pe}$ due to the approximation error introduced by Jensen's inequality in Eq.~\eqref{eq:KawasakiCrooksVariationalJensen}.
The control potentials used to evaluate $\hat{F}^\mathrm{neq}_\mathrm{VTR}(r)$ are shown in Figs.~\ref{fig:sheared_particle_control_policy}A-D.
Each potential was estimated by solving the velocity matching problem in Eq.~\eqref{eq:FokkerPlanckVelocityMSE} with the control policy ansatz in Eq.~\eqref{eq:FokkerPlanckVelocityAnsatz} and
\begin{equation}\label{eq:ShearedParticleControlPotentialAnsatz}
\begin{gathered}
  V_{\boldsymbol{\theta}}(X) \,=\!
    \sum\limits_{i_{x_1}=\,0}^{N_{x_1}\!-1}\sum\limits_{i_{x_2}=\,0}^{N_{x_2}\!-1}%
    \theta_{i_{x_1}, i_{x_2}} \exp \, \bigl( -s^2 |X - \mu_{i_{x_1}, i_{x_2}}|^2 /\:\! 2 \bigr)
  \\%
  \mu_{i_{x_1}, i_{x_2}} \!=\, \Delta\mu \, \bigl(
    i_{x_1} - (N_{x_1}-1)/2,\, i_{x_2} - (N_{x_2}-1)/2%
  \bigr)^\top
\end{gathered}
\end{equation}
where the parameter vector $\boldsymbol{\theta} \equiv (\theta_{i_{x_1}, i_{x_2}})$ holds amplitude coefficients for squared-exponential functions with scale $s$ and locations $\bm\{ \mu_{i_{x_1}, i_{x_2}} \bm\}$ spanning a uniform grid on the $x_1x_2$-plane.
For all $\mathrm{Pe}$, the control potential ansatz was instantiated using $256$ squared-exponential basis functions with scale parameter $s = 0.25$ and locations spanning a $16 \times 16$ origin-centered grid with spacing $\Delta \mu = 0.25$ along the $x_1$ and $x_2$ coordinates.
Optimal controller amplitude coefficients for the basis functions were obtained by solving for the parameter vector $\boldsymbol{\theta}$ at the global minimum of the functional in Eq.~\eqref{eq:FokkerPlanckVelocityMSE}, using the system of equations derived for linear controller ans\"{a}tze in Appendix~\ref{sec:AppendixC}.
At each shear rate, the steady-state averages used to solve for the optimal controller were estimated from $512$ independent steady-state realizations of  Eq.~\eqref{eq:NonequilibriumSDE} at least $25$ time units long.
 
The optimized control potentials in Figs.~\ref{fig:sheared_particle_control_policy}A-D show how particle configurations near the barrier top are stabilized by the nonequilibrium shear, thereby reducing the effective free energy barrier along the $x_1$ coordinate with increasing shear rate, as observed in Figs.~\ref{fig:sheared_particle_schematic_and_feps}B-E.
Moreover, the sum of the equilibrium potential $U$ and the control potential $V_{\boldsymbol{\theta}}$, shown in Fig.~\ref{fig:sheared_particle_control_policy}E at $\mathrm{Pe} = 16$, yields an effective equilibrium system with the correct nonequilibrium statistics of the sheared particle system, from which the nonequilibrium free energy surface in Fig.~\ref{fig:sheared_particle_schematic_and_feps}E can be recovered using equilibrium FES estimation methods.

\begin{figure}[t]
\centering
\includegraphics[width=0.95\columnwidth]%
  {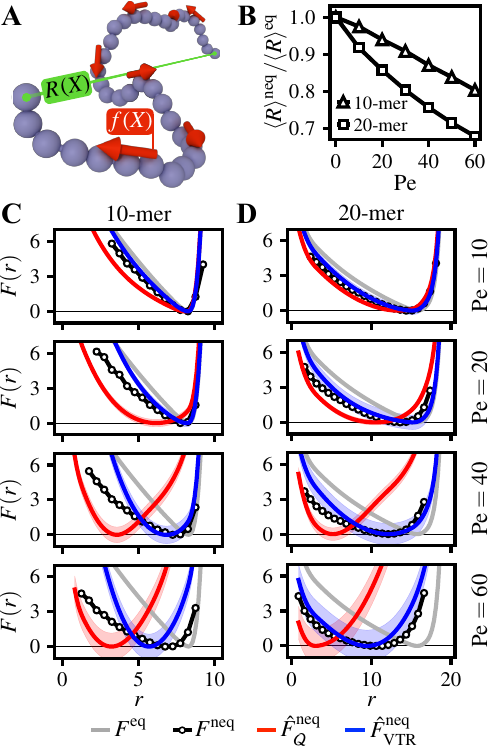}
\caption{%
  \textbf{A}.
  Schematic of a $20$-mer active filament depicting the swimming force $f(X)$ [Eq.~\eqref{eq:SwimmingFilamentNonequilibriumDrift}] and the end-to-end distance coordinate $R(X)$.
  \textbf{B}.
  Mean normalized end-to-end distance $\langle R \rangle^\mathrm{neq} / \langle R \rangle^\mathrm{eq}$ as a function of swimming speed $\mathrm{Pe}$ for $10$-mer and $20$-mer filaments.
  \textbf{C}, \textbf{D}.
  Free energy surfaces along the end-to-end distance coordinate at various $\mathrm{Pe}$ based on several nonequilibrium estimators (colored lines), together with the equilibrium free energy surface $F^\mathrm{eq}(r)$ (gray line).
  The black circles estimate the nonequilibrium free energy surface $F^\mathrm{neq}(r)$ from histogram data at each $\mathrm{Pe}$.
  The shaded regions surrounding the free energy estimates indicate their statistical uncertainty, at approximately one standard deviation from the mean.
}\label{fig:swimming_filament_schematic_and_feps}
\end{figure}

\begin{figure*}[t]
\centering
\includegraphics[width=\textwidth]%
  {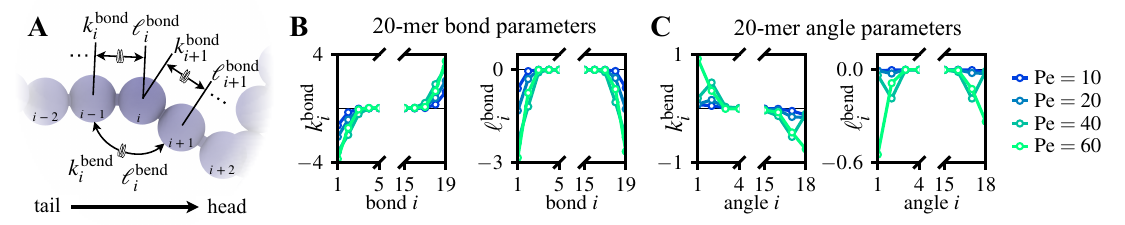}
\caption{%
  \textbf{A}.
  Schematic of parameters in the control policy ansatz, defined in Eq.~\eqref{eq:SwimmingFilamentControlPotentialAnsatz}.
  \textbf{B}, \textbf{C}.
  Dimensionless optimized values for control policy parameters for $20$-mer at various $\mathrm{Pe}$.
  Bond and angle parameter values close to the swimming filament's termini are respectively shown in~B and~C.
  Monomers, bonds, and angles along the filament are indexed counting from the tail terminus to the head terminus, in increasing order along the swimming direction.
}\label{fig:swimming_filament_control_policy}
\end{figure*}

\subsection{Swimming filament}%
\label{ssec:SwimmingFilament}

As a more challenging test of the nonequilibrium FES 
estimation methods introduced in Section~\ref{sec:TheoryAndMethod}, we estimate the nonequilibrium FES 
along the end-to-end distance coordinate of the $N$-monomer swimming filament depicted in Fig.~\ref{fig:swimming_filament_schematic_and_feps}A.
The filament's dynamics is governed by the SDE in Eq.~\eqref{eq:NonequilibriumSDE} with $U(X)$ a discretization of the potential energy of a self-excluding semiflexible Gaussian chain~\cite{doi1988theory} and $f(X)$ a swimming force tangential to the filament.
Representing the filament configuration as $X \!=\! (X^{(j)})_{j=1}^N$, with $X^{(j)}$ the position of the $j$th monomer along the chain, the potential energy comprises bond, bend, and WCA-type~\cite{weeks1971role} exclusion interactions as
\begin{equation}\label{eq:SwimmingFilamentPES}
  U(X) \,=\, U_\mathrm{bond}(X) \,+\, U_\mathrm{bend}(X) \,+\, U_\mathrm{excl}(X)
\end{equation}
where
\begin{equation*}
\begin{alignedat}{3}
  U_\mathrm{bond}(X) \,&=\, \dfrac{k_\mathrm{bond}}{2} && \textstyle\sum_{j=2}^{N} ( |X^{(j)} - X^{(j-1)}| - \ell_\mathrm{bond} )^2 \\[4pt]
  U_\mathrm{bend}(X) \,&=\, \dfrac{k_\mathrm{bend}}{2} && \textstyle\sum_{j=3}^{N} |X^{(j)} - 2X^{(j-1)} + X^{(j-2)}|^2 \\
  U_\mathrm{excl}(X) \,&=\, 2\varepsilon_\mathrm{excl} && \textstyle\sum_{\substack{j \neq k = 1}}^{N} \bm{\biggl[} \dfrac{1}{4} + \biggl( \dfrac{\sigma_\mathrm{excl}}{d_{j,k}} \biggr)^{\!12} - \biggl( \dfrac{\sigma_\mathrm{excl}}{d_{j,k}} \biggr)^{\!6} \, \bm{\biggr]}
\end{alignedat}
\end{equation*}
with $d_{j,k} \!=\! \min(|X^{(j)} - X^{(k)}|,\, 2^{1/6} \sigma_\mathrm{excl})$.
We let $\varepsilon_\mathrm{excl} \equiv 1$ and $\sigma_\mathrm{excl} \equiv 1$, and set $k_\mathrm{bond} \!=\! 1000 (\beta\sigma_\mathrm{excl}^{2})^{-1}$, $k_\mathrm{bend} \!=\! 10 (\beta\sigma_\mathrm{excl}^{2})^{-1}$, and $\ell_\mathrm{bond} \!=\! 2^{1/6} \sigma_\mathrm{excl}$.
The filament swims along its contour at speed $\mathrm{Pe}$, advected with the nonequilibrium drift $f(X) = \bm( f^{(j)}(X) \bm)_{j=1}^{N}$ where
\begin{equation}\label{eq:SwimmingFilamentNonequilibriumDrift}
  f^{(j)}(X) \,=\,
  \dfrac{\mathrm{Pe}}{2} \times
  \begin{cases}
    U_{j,j+1} & j=1 \\
    U_{j,j+1} + U_{j-1,j} & j=2,\!...,N\!-1 \\
    U_{j,j-1} & j=N \\
  \end{cases}
\end{equation}
is the swimming force on the $j$th monomer, defined in terms of $U_{j,k} \!=\! (X^{(k)} - X^{(j)}) / |X^{(k)} - X^{(j)}|$ the unit vector parallel to the displacement between monomers $j$ and $k$.

The phenomenology of the swimming filament has been explored across a broad range of monomer numbers $N$ in earlier work~\cite{Bianco2018,Anand2018}, and is illustrated for $10$-mer and $20$-mer filaments in Fig.~\ref{fig:swimming_filament_schematic_and_feps}B.
Notably, the steady-state mean end-to-end distance $R(X) = |X^{(N)} - X^{(1)}|$ monotonically decreases with increasing swimming speed, since the head of the swimming filament swerves more frequently at higher swimming speeds. This swerve of the head, which moves more slowly than the body of the filament, results in a buckling of filament configurations.
These buckling motions become more typical as the system is driven away from equilibrium.
Further, larger chains have large propulsive forces, as they are additive along the chain, resulting in higher likelihood of buckling and a more pronounced decrease of end-to-end distance with $\mathrm{Pe}$.

To estimate the FES along the end-to-end distance coordinate, $r$-conditioned equilibrium ensembles were approximated with umbrella-biased equilibrium ensembles per
\begin{equation*}
  \rho^\mathrm{eq}_r(X) \,\approx\, \rho^\mathrm{eq}(X) \, \dfrac{\exp \, \bm[ -\beta U_\mathrm{umb} \bm( R(X) \bm) \bm]}{\textstyle\int_{\Omega} \mathrm{d}X' \, \exp \, \bm[ -\beta U_\mathrm{umb} \bm( R(X') \bm) \bm]}
\end{equation*}
where the umbrella potentials
\begin{equation*}
  U_\mathrm{umb}(r) \,=\, k_\mathrm{umb} (r - \ell_\mathrm{umb})^2 /\:\! 2
\end{equation*}
were assigned stiffness $k_\mathrm{umb} = 100(\beta\sigma_\mathrm{excl}^{2})^{-1}$ and locations $\ell_\mathrm{umb}$ spanning a uniform grid along the end-to-end distance coordinate, with values inside the interval $\bm{\bigl(} 0, (N-1) \ell_\mathrm{bond} \bm{\bigr)}$ for each $N$-mer.
The WHAM procedure~\cite{Kumar1992,Roux1995} was used to estimate the equilibrium FES $F^\mathrm{eq}(r)$ from the array of umbrella-biased ensembles.
Separately, a reference estimate of the nonequilibrium FES $F^\mathrm{neq}(r)$ at swimming speeds $\mathrm{Pe} \in \{10, 20, 40, 60\}$ was obtained from a histogram of the end-to-end distance along a steady-state realization of the SDE in Eq.~\eqref{eq:NonequilibriumSDE} integrated out to $10^3$ time units using the Euler--Maruyama method, with stepsizes as small as $10^{-5}$ for numerical stability.
Nonequilibrium FES estimates $\hat{F}^\mathrm{neq}_{\mathcal{Q}}(r)$ and $\hat{F}^\mathrm{neq}_{\mathrm{VTR}}(r)$ were evaluated by integrating $10^5$ relaxation trajectories using the SDEs in Eqs.~\eqref{eq:NonequilibriumSDE} and~\eqref{eq:ControlledNonequilibriumSDE}, respectively, with the conformation $r$ of each estimate given by the mean end-to-end distance of the approximately conditioned ensemble in which relaxation trajectories were initialized.
All relaxation trajectories were integrated out to $25$ time units using the Euler--Maruyama method at stepsizes as small as $10^{-6}$ time units to ensure numerical stability.

FES estimates from steady-state trajectory histogramming and from variational estimators are shown for $10$-mer and $20$-mer filaments in Figs.~\ref{fig:swimming_filament_schematic_and_feps}C and~\ref{fig:swimming_filament_schematic_and_feps}D, for swimming speeds ranging between $\mathrm{Pe} = 10$ and $\mathrm{Pe} = 60$ units.
The equilibrium free energy surface for each $N$-mer is also shown in its respective panels. 
Across all $\mathrm{Pe}$ values for the $10$-mer and $20$-mer filaments, we find better qualitative agreement between the brute-force nonequilibrium free energy and the variational estimator $\hat{F}^\mathrm{neq}_{\mathrm{VTR}}(r)$ than the perturbative $\hat{F}^\mathrm{neq}_{\mathcal{Q}}(r)$, though both estimators exhibit notable bias at higher $\mathrm{Pe}$ values.
Both estimators correctly predict a collapse of the filament with increasing $\mathrm{Pe}$, though $\hat{F}^\mathrm{neq}_{\mathcal{Q}}(r)$ underestimates the value of $\mathrm{Pe}$ where this occurs, indicated by the minimum in $F^\mathrm{neq}(r)$, which shifts toward smaller values of $r$ with increasing $\mathrm{Pe}$.
The variational estimator accurately predicts the behavior of $F^\mathrm{neq}(r)$ near typical values of $r$ across the range of $\mathrm{Pe}$ for both chains, but exhibits deviation from the ground truth at atypical values of $r$ for the $10$-mer. 
The traffic-based perturbative estimator $\hat{F}^\mathrm{neq}_{\mathcal{T}}(r)$ is sufficiently inaccurate for all $\mathrm{Pe}$ as to be remotely outside the range in Figs.~\ref{fig:swimming_filament_schematic_and_feps}C and~\ref{fig:swimming_filament_schematic_and_feps}D, and is hence omitted in the current discussion.

Figure~\ref{fig:swimming_filament_control_policy}A illustrates the interactions that comprise the control potential used to evaluate the nonequilibrium FES estimator $\hat{F}^\mathrm{neq}_{\mathrm{VTR}}(r)$.
Control potential terms are similar to the bond and bend interactions in the potential energy function in Eq.~\eqref{eq:SwimmingFilamentPES}, but have individual stiffness and scale parameters for each bond and bend interaction along the filament.
The control potential entering $\varv_{\boldsymbol{\theta}}$ is
\begin{equation}\label{eq:SwimmingFilamentControlPotentialAnsatz}
\begin{aligned}
  & V_{\boldsymbol{\theta}}(X) =\,%
    \textstyle\sum_{j=1}^{N-1}%
      k_j^\mathrm{bond} (|X^{(j+1)} - X^{(j)}| - \ell_j^\mathrm{bond})^2 /\:\! 2 \\%
  & \;\,+\,%
    \textstyle\sum_{j=1}^{N-2}%
      k_j^\mathrm{bend} (|X^{(j+2)} - 2X^{(j+1)} + X^{(j)}| - \ell_j^\mathrm{bend})^2 /\:\! 2
\end{aligned}
\end{equation}
where $\boldsymbol{\theta} \equiv \{ (k_j^\mathrm{bond}, \ell_j^\mathrm{bond})_{j=1}^{N-1}, (k_j^\mathrm{bend}, \ell_j^\mathrm{bend})_{j=1}^{N-2} \}$ is the policy parameter vector.
Control policy parameters were optimized at each swimming speed $\mathrm{Pe}\!\in\!\{10, 20, 40, 60\}$ and each monomer number $N\!\in\! \{10, 20\}$ via stochastic gradient descent of the loss function in Eq.~\eqref{eq:control_policy_loss} in Appendix~\ref{sec:AppendixC}, using $128$ independent steady-state realizations of the nonequilibrium SDE in Eq.~\eqref{eq:NonequilibriumSDE} at least $25$ time units long, and learning rates up to $10^{-1}$.
Optimized control potential parameters for a swimming $20$-mer at speeds ranging from $\mathrm{Pe} = 10$ to $\mathrm{Pe} = 60$ units are shown in Fig.~\ref{fig:swimming_filament_control_policy}B and Fig.~\ref{fig:swimming_filament_control_policy}C, which respectively show bond and bend interaction parameters near the swimming filament's endpoints.
Viewed as a correction to the filament's potential energy function, the optimized control potential shifts internal tensions within the swimming filament to stabilize emergent bond and bending deformations near the filament's tail ($1$st) and head ($N$th) monomers.
The localization of the optimized control potential to the boundaries of the chain explains the higher accuracy of the results for the $20$-mer compared to the $10$-mer in Figs.~\ref{fig:swimming_filament_schematic_and_feps}C and~\ref{fig:swimming_filament_schematic_and_feps}D.
For longer chains, the potential contribution of the control policy plays a smaller role than the reversing of the active force, and so comparable errors in the estimate of the optimal control in the larger chain result in small biases to the free energy surface. 

\section{Conclusions}%
\label{sec:Conclusions}

Growing interest in the physics of driven and active matter demands rapid progress toward turnkey simulation methods for the thermodynamic modeling of structure and stability of systems driven far from equilibrium.
Building from a natural generalization of the thermodynamic free energy surface to nonequilibrium systems, the current work adopts an optimal control viewpoint of a key result in stochastic thermodynamics---the Kawasaki--Crooks relation---to offer a free-energy surface estimation scheme for nonequilibrium steady states governed by overdamped Langevin dynamics.
Illustrative applications of the method to models of driven and active matter show that approximately optimally controlled nonequilibrium free energy estimators can exhibit reduced bias relative to contemporary perturbative approaches.
The control-based free energy estimator does require optimizing a control policy ansatz in each application, but as shown above, simple policy parametrizations can be optimized with little computational overhead and in an offline manner through regression from steady-state trajectory data.

Our formal approach in this work highlights deep connections between the fluctuation theorems at the heart of stochastic thermodynamics and the formalism of stochastic control theory~\cite{Anderson1982,Pavon1989,Pra1991,Chetrite2011}, paving the way to migrate generative modeling tools based on variational time reversal of diffusive processes~\cite{Sohl-Dickstein2015,Song2021amended,DeBortoli2021} into chemical and biological physics for nonequilibrium structure and stability quantification.

All numerical results presented in this work, along with generating code, are deposited in the GitHub repository~\href{https://github.com/jrosaraices/nonequilibrium-vtr}{\texttt{jrosaraices/nonequilibrium-vtr}}.

\begin{acknowledgments}
This work was supported by the U.S. Department of Energy, Office of Science, Office of Advanced Scientific Computing Research, and Office of Basic Energy Sciences, via the Scientific Discovery through Advanced Computing (SciDAC) program.
J.L.R.-R. acknowledges support from the National Science Foundation through the MPS-Ascend Postdoctoral Research Fellowship, Award No.\ 2213064.
We thank Aditya Singh, Adrianne Zhong, and Gavin Crooks for fruitful conversations.
\end{acknowledgments}

\appendix

\section{Derivation of the Kawasaki--Crooks relation for nonequilibrium free energy shifts}%
\label{sec:AppendixA}

In this Appendix, we derive the Kawasaki--Crooks equality in Eq.~\eqref{eq:KawasakiCrooks}, relating the free energy surface shift in Eq.~\eqref{eq:FESurfaceShift} to an average over nonequilibrium trajectories initialized at conformation $r$.
To this end, we consider the Dirac observable $\delta\bm(R(X_T) - r\bm)$ that yields the time-dependent nonequilibrium likelihood at $r$ through the relation $P_T(r) = \langle \delta\bm(R(X_T) - r\bm)\rangle^\mathrm{neq}$, where the superscript on the angled brackets denotes expectation with respect to the path probability measure induced by the SDE in Eq.~\eqref{eq:NonequilibriumSDE} with equilibrium initial conditions.
This expectation may be manipulated as
\begingroup
\allowdisplaybreaks%
\begin{subequations}\label{eq:KawasakiCrooksDerivation}
\begin{align}
  P_T(r)
  \, &= \,
  \langle \delta\bm( R(X_T) - r \bm) \rangle^\mathrm{neq}%
  \label{eq:KawasakiCrooksDerivation_1} \\
  \, &= \,
  \langle \exp \, \bm[ -\beta \mathcal{S}_T(\boldsymbol{X}) \bm] \,\delta\bm( R(X_T) - r \bm) \rangle^\mathrm{eq}%
  \label{eq:KawasakiCrooksDerivation_2} \\
  \, &= \,
  \langle \exp \, \bm[ -\beta \mathcal{S}_T(\tilde{\boldsymbol{X}}) \bm] \,\delta\bm( R(X_0) - r \bm) \rangle^\mathrm{eq}%
  \label{eq:KawasakiCrooksDerivation_3} \\
  \, &= \,
  \langle \exp \, \bm[ -\beta \mathcal{S}_T(\tilde{\boldsymbol{X}}) \bm] \rangle_r^\mathrm{eq} \, P_0(r)%
  \label{eq:KawasakiCrooksDerivation_4} \\
  \, &= \,
  \langle \exp \, \bm[ -\beta (\mathcal{S}_T - \mathcal{Q}_T)(\boldsymbol{X}) \bm] \rangle_r^\mathrm{eq} \, P_0(r)%
  \label{eq:KawasakiCrooksDerivation_5} \\
  \, &= \,
  \langle \exp \, \bm[ \beta \mathcal{Q}_T(\boldsymbol{X}) \bm] \rangle_r^\mathrm{neq} \, P_0(r)
  \label{eq:KawasakiCrooksDerivation_6}
\end{align}
\end{subequations}
\endgroup
In Eq.~\eqref{eq:KawasakiCrooksDerivation_2}, the path expectation in Eq.~\eqref{eq:KawasakiCrooksDerivation_1} is rewritten in terms of an equilibrium path ensemble generated by the SDE in Eq.~\eqref{eq:EquilibriumSDE} initialized at equilibrium, with the pathwise change of measure given by Eq.~\eqref{eq:RelativeDensity}.
Invariance under time reversal of the equilibrium path ensemble allows shifting the Dirac observable across the time interval via pathwise time involution, which yields Eq.~\eqref{eq:KawasakiCrooksDerivation_3}.
Then, the Markov property allows introducing the conditional expectation in Eq.~\eqref{eq:KawasakiCrooksDerivation_4}, which averages over paths initialized per the restricted equilibrium density in Eq.~\eqref{eq:RestrictedEquilibriumPDF}.
Finally, Eq.~\eqref{eq:RelativeDensityTimeInvolution} is used to write the change of measure in terms of the excess heat in Eq.~\eqref{eq:KawasakiCrooksDerivation_5}, and the remainder is absorbed into the path measure to reintroduce the nonequilibrium path ensemble in Eq.~\eqref{eq:KawasakiCrooksDerivation_6}.

From Eq.~\eqref{eq:KawasakiCrooksDerivation}, the nonequilibrium free energy shift in the long-time limit takes the form
\begin{equation}
\begin{aligned}
  \ln \dfrac{P^\mathrm{neq}(r)}{P^\mathrm{eq}(r)}
  \,&=\,
  \lim_{T \to \infty} \ln \dfrac{P_T(r)}{P_0(r)} \\
  \,&=\,
  \lim_{T \to \infty} \ln
  {\langle \exp \, \bm[ \beta \mathcal{Q}_T(\boldsymbol{X}) \bm] \rangle^\mathrm{neq}_r}%
\end{aligned}
\end{equation}
which corresponds to the sought-after result.

The Kawasaki--Crooks equality in Eq.~\eqref{eq:KawasakiCrooks} is related to the integral fluctuation relation in Eq.~\eqref{eq:IFT} for the excess heat $\mathcal{Q}_T$.
This is a consequence of the fact that
\begin{subequations}\label{eq:IFTDerivation}
\begin{align}
  \langle \exp \, \bm[ \beta \mathcal{Q}_T(\boldsymbol{X}) \bm] \rangle^\mathrm{neq}
  \,&=\,
  \langle \exp \, \bm[ -\beta (\mathcal{S}_T - \mathcal{Q}_T)(\boldsymbol{X}) \bm] \rangle^\mathrm{eq}
  \label{eq:IFTDerivation_1} \\
  \,&=\,
  \langle \exp \, \bm[ -\beta \mathcal{S}_T(\tilde{\boldsymbol{X}}) \bm] \rangle^\mathrm{eq}
  \label{eq:IFTDerivation_2} \\
  \,&=\,
  \langle \exp \, \bm[ -\beta \mathcal{S}_T(\boldsymbol{X}) \bm] \rangle^\mathrm{eq}
  \label{eq:IFTDerivation_3} \\
  \,&=\,
  \langle 1 \rangle^\mathrm{neq} = 1
  \label{eq:IFTDerivation_4}
\end{align}
\end{subequations}
where Eq.~\eqref{eq:IFTDerivation_3} used the fact that the path ensemble generated by the equilibrium SDE in Eq.~\eqref{eq:EquilibriumSDE} with equilibrium initial conditions is invariant under time reversal.

\section{Derivation and evaluation of the optimal control policy for a configuration-wise Kawasaki--Crooks equality}%
\label{sec:AppendixB}

For the identity order parameter such that $R(X) = X$ specifies an initial \emph{configuration} (rather than an equilibrium ensemble of configurations with a given \emph{conformation} $r$), the general Kawasaki--Crooks equality in Eq.~\eqref{eq:KawasakiCrooks} reduces to the configuration-wise form
\begin{equation}\label{eq:KawasakiCrooksPointwise}
  \dfrac{\rho^\mathrm{neq}(X)}{\rho^\mathrm{eq}(X)}
  \,=\, \lim_{T \to \infty} \langle \exp \, \bm[ \beta \mathcal{Q}_T(\boldsymbol{X}) \bm] \rangle_X^\mathrm{neq} 
  \,\equiv\, \exp \bm(V(X)\bm)
\end{equation}
Applying the Feynman--Ka\c{c} theorem~\cite[Theorem~10.2]{Prato2014}
to the exponential average reveals that $\exp(V)$ solves the steady-state backward Kolmogorov equation 
\begin{equation}\label{eq:KawasakiCrooksTiltedBKE}
  0 \,=\, -\bm[ \mathscr{L} - 2f \cdot \nabla + \beta \nabla U \cdot f \bm] \exp(V)
\end{equation}
where $\mathscr{L}$ is the generator of the nonequilibrium trajectory ensemble governed by the SDE in Eq.~\eqref{eq:NonequilibriumSDE}, namely
\begin{equation}
  \mathscr{L} \,=\, (f - \nabla U) \cdot \nabla + \beta^{-1} \Delta
\end{equation}
The \emph{tilted} generator in Eq.~\eqref{eq:KawasakiCrooksTiltedBKE} is associated with a non measure-preserving stochastic dynamics and would suggest that Eq.~\eqref{eq:KawasakiCrooksPointwise} must be estimated in simulation through population dynamics approaches, which are notorious for their high statistical variance in the relevant long-time limit~\cite{Ray2018a,Ray2018b,Ray2020,Angeli2021}.
Fortunately, the generalized Doob transform~\cite{Chetrite2014}
of the tilted generator yields a statistically equivalent, yet measure-preserving, stochastic dynamics that allows minimal-variance estimation of the exponential averages in Eq.~\eqref{eq:KawasakiCrooksPointwise}.

To compute the generalized Doob transform, observe that the tilted generator satisfies the operator equality
\begin{equation}\label{eq:KawasakiCrooksTiltedGenerator}
  \mathscr{L} - 2f \cdot \nabla + \beta \nabla U \cdot f
  \,=\, (\rho^\mathrm{eq})^{-1} \circ \mathscr{L}^\dag \circ \rho^\mathrm{eq}
\end{equation}
with $\mathscr{L}^\dag$ the Fokker--Planck operator introduced in Eq.~\eqref{eq:FokkerPlanckEquation}.
Indeed, for any suitably differentiable scalar function $g$, 
\begingroup
\allowdisplaybreaks%
\begin{align*}
  & \bm[ (\rho^\mathrm{eq})^{-1} \circ \mathscr{L}^\dag \circ \rho^\mathrm{eq} \bm] \, g
  \\[2pt] &\;=\,
  (\rho^\mathrm{eq})^{-1} \bm\{ -\nabla \cdot \bm[ (f - \nabla U) \, \rho^\mathrm{eq} g \bm] + \beta^{-1} \Delta (\rho^\mathrm{eq} g) \bm\}
  \\[2pt] &\;=\,
  (-f - \nabla U) \cdot \nabla g + \beta^{-1} \Delta g + \beta g \, \nabla U \cdot f
  \\[2pt] &\;=\,
  \bm[ \mathscr{L} - 2f \cdot \nabla + \beta \nabla U \cdot f \bm] \, g
\end{align*}
\endgroup
Using Eqs.~\eqref{eq:KawasakiCrooksPointwise} and~\eqref{eq:KawasakiCrooksTiltedGenerator} reveals the Doob-transformed tilted generator in Eq.~\eqref{eq:KawasakiCrooksTiltedBKE} is
\begin{equation}\label{eq:KawasakiCrooksDoobGenerator}
\begin{aligned}
  & \exp(-V) \circ \bm[ (\rho^\mathrm{eq})^{-1} \circ \mathscr{L}^\dag \circ \rho^\mathrm{eq} \bm] \circ \exp(V)
  \\ &\quad=\, (\rho^\mathrm{ne})^{-1} \circ \mathscr{L}^\dag \circ \rho^\mathrm{ne}
  \\ &\quad=\, \mathscr{L} - 2(f - \beta^{-1} \nabla V) \cdot \nabla
\end{aligned}
\end{equation}
which corresponds to the controlled SDE in Eq.~\eqref{eq:ControlledNonequilibriumSDE} with the control policy $\varv^\mathrm{neq}$ in Eq.~\eqref{eq:FokkerPlanckVelocity}.
To see that this controlled dynamics indeed yields a minimum-variance variational estimate of Eq.~\eqref{eq:KawasakiCrooksPointwise}, note that for \emph{every} realization $\boldsymbol{X}$ of the nonequilibrium SDE in Eq.~\eqref{eq:ControlledNonequilibriumSDE} initialized at the configuration $X$, the logarithm of the argument of the expectation in the second line of Eq.~\eqref{eq:KawasakiCrooksVariational} reduces to
\begin{widetext}
\begin{subequations}\label{eq:KawasakiCrooksPointwiseVariationalOptimumPathwise}
\begin{align}
  \beta \mathcal{Q}_T(\boldsymbol{X}) + \ln \dfrac{\mathrm{d}\mathbb{P}^\mathrm{neq}(\boldsymbol{X})}{\mathrm{d}\mathbb{P}^{\varv^\mathrm{neq}}(\boldsymbol{X})}
  \,&=\, \beta \mathcal{Q}_T(\boldsymbol{X}) - \beta \textstyle\int_0^T \!\mathrm{d}t \, |\varv^\mathrm{neq}(X_t)|^2 + \sqrt{2\beta} \textstyle\int_0^T \mathrm{d}W_t \cdot \varv^\mathrm{neq}(X_t)
  \label{eq:KawasakiCrooksPointwiseVariationalOptimumPathwise_1} \\
  \,&=\, \beta \mathcal{Q}_T(\boldsymbol{X}) + \beta \textstyle\int_0^T \!\mathrm{d}t \, \bm\{ |\varv^\mathrm{neq}(X_t)|^2 + \bm[ (\beta^{-1} \nabla + f - \nabla U) \cdot \varv^\mathrm{neq} \bm](X_t) \bm\} - \mathrm{d}X_t \circ \varv^\mathrm{neq}(X_t)
  \label{eq:KawasakiCrooksPointwiseVariationalOptimumPathwise_2} \\[4pt]
  \,&=\, \textstyle\int_0^T \!\mathrm{d}t \, \bm[ \beta^{-1} (|\nabla V|^2 + \Delta V) - (f + \nabla U) \cdot \nabla V + \beta \nabla U \cdot f \bm](X_t) - \mathrm{d}X_t \circ \nabla V(X_t)
\label{eq:KawasakiCrooksPointwiseVariationalOptimumPathwise_3} \\[4pt]
  \,&=\, \textstyle\int_0^T \!\mathrm{d}t \exp\bm(-V(X_t)\bm) \, \bm[ \mathscr{L} - 2f \cdot \nabla + \beta \nabla U \cdot f \bm] \exp\bm(V(X_t)\bm) - \mathrm{d}X_t \circ \nabla V(X_t)
\label{eq:KawasakiCrooksPointwiseVariationalOptimumPathwise_4} \\[4pt]
  \,&=\, -\textstyle\int_0^T \mathrm{d}X_t \circ \nabla V(X_t) \,=\, V(X) - V(X_T)
\label{eq:KawasakiCrooksPointwiseVariationalOptimumPathwise_5}
\end{align}
\end{subequations}
\end{widetext}
where It\={o}'s formula~\cite[Chapter~3.4]{Pavliotis2014} obtains
Eq.~\eqref{eq:KawasakiCrooksPointwiseVariationalOptimumPathwise_2}, the heat functional vanished due to cancellation in
Eq.~\eqref{eq:KawasakiCrooksPointwiseVariationalOptimumPathwise_3}, the Hopf--Cole substitution $V \!\to \exp(V)$ leads to
Eq.~\eqref{eq:KawasakiCrooksPointwiseVariationalOptimumPathwise_4}, and Eq.~\eqref{eq:KawasakiCrooksTiltedBKE} finally yields the potential difference
Eq.~\eqref{eq:KawasakiCrooksPointwiseVariationalOptimumPathwise_5}.
Averaging this difference against realizations of the optimally controlled SDE in Eq.~\eqref{eq:ControlledNonequilibriumSDE} yields a remarkably simple lower bound on the configuration-wise Kawasaki--Crooks equality in Eq.~\eqref{eq:KawasakiCrooksPointwise}:
\begin{equation}\label{eq:KawasakiCrooksPointwiseVariationalOptimum}
\begin{aligned}
  \ln \dfrac{\rho^\mathrm{neq}(X)}{\rho^\mathrm{eq}(X)}
  \,&\ge\, \lim_{T \to \infty} \Bigl\langle%
    \beta \mathcal{Q}_T(\boldsymbol{X}) + \ln \dfrac{\mathrm{d}\mathbb{P}^\mathrm{neq}(\boldsymbol{X})}{\mathrm{d}\mathbb{P}^{\varv^\mathrm{neq}}(\boldsymbol{X})}%
  \Bigr\rangle_X^{\varv^\mathrm{neq}}%
  \\[2pt]
  &=\, V(X) - \lim_{T \to \infty} \langle V(X_T) \rangle_X^{\varv^\mathrm{neq}}%
  \\
  &=\, \ln \dfrac{\rho^\mathrm{neq}(X)}{\rho^\mathrm{eq}(X)} - \mathcal{D}_\mathrm{KL}( \rho^\mathrm{neq} \Vert \rho^\mathrm{eq} )
\end{aligned}
\end{equation}
This lower bound on Eq.~\eqref{eq:KawasakiCrooksPointwise} is free of trajectory functionals associated with time-extensive statistical variance, instead restating the trivial inequality for the scalar nonequilibrium free energy of the NESS in Eq.~\eqref{eq:NonequilibriumFE}.
The Kullback--Leibler divergence appears in the last equality in Eq.~\eqref{eq:KawasakiCrooksPointwiseVariationalOptimum} because the controlled SDE in Eq.~\eqref{eq:ControlledNonequilibriumSDE}, when equipped with the control policy $\varv^{\boldsymbol{\theta}} = \varv^\mathrm{neq}$, has the same invariant distribution as the SDE in Eq.~\eqref{eq:NonequilibriumSDE}.
This follows from the fact that the $\mathrm{L}^2(\Omega)$-adjoint of the generator in Eq.~\eqref{eq:KawasakiCrooksDoobGenerator} also satisfies the steady-state Fokker--Planck equation in Eq.~\eqref{eq:FokkerPlanckEquation}.
Hence, an unbiased estimate of the Kullback--Leibler divergence in Eq.~\eqref{eq:KawasakiCrooksPointwiseVariationalOptimum} can be obtained from a time average of the optimal control potential $V$ over a long realization of either the nonequilibrium SDE in Eq.~\eqref{eq:NonequilibriumSDE} or the controlled SDE in Eq.~\eqref{eq:ControlledNonequilibriumSDE} with control policy $\varv^{\boldsymbol{\theta}} = \varv^\mathrm{neq}$.

\section{Derivation of a linear solution to the Fokker--Planck velocity matching problem}%
\label{sec:AppendixC}

As discussed in Sec.~\ref{ssec:TheoryAndMethod3} of the main text, the optimal control policy in Eq.~\eqref{eq:FokkerPlanckVelocity} must be estimated to evaluate the Kawasaki--Crooks estimator $\hat{F}^\mathrm{neq}_\mathrm{VTR}(r)$ in Eq.~\eqref{eq:KawasakiCrooksVariationalJensen}.
This entails minimizing the mean-squared error function in Eq.~\eqref{eq:FokkerPlanckVelocityMSE}, namely
\begin{equation*}
  \lim_{T \to \infty} \tfrac{1}{T} \Bigl\langle \textstyle\int_0^T \!\mathrm{d}t \, |\varv^\mathrm{neq}(X_t) - \varv_{\boldsymbol{\theta}}(X_t)|^2 \Bigr\rangle^\mathrm{neq}
\end{equation*}
with respect to policy parameters $\boldsymbol{\theta}$.
Expanding the square inside the expectation produces linear and quadratic terms in the unknown target policy $\varv^\mathrm{neq}$, which must be eliminated for actual computation.
The quadratic term can simply be dropped, and for the linear term we observe that for any suitably differentiable vector function $h$,
\begingroup
\allowdisplaybreaks%
\begin{align*}
  \Bigl\langle &\textstyle\int_0^T \!\mathrm{d}t \,
  h(X_t) \cdot \varv^\mathrm{neq}(X_t)
  \Bigr\rangle^\mathrm{neq}
  \\[4pt] &=
  \Bigl\langle \textstyle\int_0^T \!\mathrm{d}t \,
    h(X_t) \cdot (f - \nabla U - \beta^{-1}\nabla \ln \rho^\mathrm{neq})(X_t)
  \Bigr\rangle^\mathrm{neq}
  \\[4pt] &=
  \Bigl\langle \textstyle\int_0^T
    h(X_t) \cdot \bm[ (f - \nabla U)(X_t) \, \mathrm{d}t + \sqrt{2\beta^{-1}} \, \mathrm{d}W_t \bm]
  \Bigr\rangle^\mathrm{neq}
  \\[4pt] &\phantom{=} -\beta^{-1} \textstyle\int_0^T \mathrm{d}t \textstyle\int_{\Omega} \mathrm{d}X \, \rho^\mathrm{neq}(X) \bm[ h(X) \cdot \nabla \ln \rho^\mathrm{neq}(X) \bm]
  \\[4pt] &=
  \Bigl\langle \textstyle\int_0^T
    h(X_t) \cdot \mathrm{d}X_t
  \Bigr\rangle^\mathrm{neq}
  \\[4pt] &\phantom{=} +\beta^{-1} \textstyle\int_0^T \mathrm{d}t \textstyle\int_{\Omega} \mathrm{d}X \, \rho^\mathrm{neq}(X) \bm( \nabla \cdot h \bm)(X)
  \\[4pt] &=
  \Bigl\langle \textstyle\int_0^T
    h(X_t) \circ \mathrm{d}X_t
  \Bigr\rangle^\mathrm{neq}
\end{align*}
\endgroup
where we used Eq.~\eqref{eq:FokkerPlanckVelocity} in the first equality, Eq.~\eqref{eq:NonequilibriumSDE} in the second equality, integration by parts in the third equality, and the definition of the Stratonovich integral in the fourth equality.

Fully eliminating the target velocity from the mean squared error yields the loss function 
\begin{equation}\label{eq:control_policy_loss}
  \mathsf{L}(\boldsymbol{\theta})
  \,=\, \lim_{T \to \infty} \tfrac{1}{T} \Bigl\langle \textstyle\int_0^T%
    |\varv_{\boldsymbol{\theta}}(X_t)|^2 \, \mathrm{d}t - 2 \varv_{\boldsymbol{\theta}}(X_t) \circ \mathrm{d}X_t%
  \Bigr\rangle^\mathrm{neq}
\end{equation}
to be minimized with respect to variations in the policy parameters $\boldsymbol{\theta}$.
Using the control policy ansatz $\varv_{\boldsymbol{\theta}}$ in Eq.~\eqref{eq:FokkerPlanckVelocityAnsatz} leaves only the control potential $V_{\boldsymbol{\theta}}$ up to parametric estimation.
The gradient of the loss function in Eq.~\eqref{eq:control_policy_loss} may be written as
\begingroup
\allowdisplaybreaks%
\begin{align*}
  \nabla_{\boldsymbol{\theta}} \mathsf{L}(\boldsymbol{\theta})
    &\,=\, \lim_{T \to \infty} \tfrac{1}{T} \Bigl\langle \textstyle\int_0^T \mathrm{d}t \, \nabla_{\boldsymbol{\theta}} \, \bm[ |\varv_{\boldsymbol{\theta}}(X_t)|^2%
  \\&\qquad\qquad\quad - 2\, \mathrm{d}X_t \circ \varv_{\boldsymbol{\theta}}(X_t) \bm] \Bigr\rangle^\mathrm{neq}%
  \\[1pt]&\,=\, \lim_{T \to \infty} \tfrac{2}{T} \Bigl\langle \textstyle\int_0^T \mathrm{d}t \, (\nabla V_{\boldsymbol{\theta}} - f)(X_t) \cdot \nabla_{\boldsymbol{\theta}} \nabla V_{\boldsymbol{\theta}}(X_t)%
  \\&\qquad\qquad\quad + \mathrm{d}X_t \circ \nabla_{\boldsymbol{\theta}} \nabla V_{\boldsymbol{\theta}}(X_t) \Bigr\rangle^\mathrm{neq}%
\end{align*}
\endgroup

A linear representation of the control potential such as
\begin{equation}\label{eq:control_potential_ansatz}
  V_{\boldsymbol{\theta}}(X) \,=\, \textstyle\sum_{i} \theta_i \phi_i(X)
\end{equation}
where $\bm\{ \phi_i(\cdot) \bm\}$ is a basis of suitably differentiable scalar functions, leads to the following system of equations for the optimal parameter vector $\boldsymbol{\theta}^\star \equiv (\theta_i^\star)$ at the global minimum of Eq.~\eqref{eq:control_policy_loss} and hence the velocity matching loss in Eq.~\eqref{eq:FokkerPlanckVelocityMSE}:
\begin{equation}\label{eq:control_potential_loss_gradient_zero}
\begin{aligned}
  0 \,=\, \lim_{T \to \infty} \tfrac{1}{T} \Bigl\langle%
    &\textstyle\int_0^T \mathrm{d}t \, \bm{\bigl[} \textstyle\sum_j \theta_j^\star \, (\nabla \phi_j - f)(X_t) \bm{\bigr]} \cdot \nabla \phi_i(X_t)%
  \\&+\,\textstyle\int_0^T \mathrm{d}X_t \circ \nabla \phi_i(X_t)%
  \Bigr\rangle^\mathrm{neq} \quad \forall \; i
\end{aligned}
\end{equation}
Building this linear system merely requires the evaluation of steady-state averages of scalar products of the differentiated basis $\bm\{ \nabla \phi_i(\cdot) \bm\}$.
In detail, Eq.~\eqref{eq:control_potential_loss_gradient_zero} may be rewritten as
\begin{equation*}
  \textstyle\sum_j \theta_j^\star \, \mathsf{A}_{ij}  \,=\, \mathsf{b}_i -\, \mathsf{c}_i \quad \forall \; i
\end{equation*}
where
\begin{equation*}
\begin{aligned}
  \mathsf{A}_{ij} \,&=\, \lim_{T \to \infty} \tfrac{1}{T} \Bigl\langle \textstyle\int_0^T%
    \mathrm{d}t \, \nabla \phi_i(X_t) \cdot \nabla \phi_j(X_t)%
  \Bigr\rangle^\mathrm{neq}
  \\[4pt]
  \mathsf{b}_i    \,&=\, \lim_{T \to \infty} \tfrac{1}{T} \Bigl\langle \textstyle\int_0^T%
    \mathrm{d}t \, \nabla \phi_i(X_t) \cdot f(X_t)%
  \Bigr\rangle^\mathrm{neq}
  \\[4pt]
  \mathsf{c}_i    \,&=\, \lim_{T \to \infty} \tfrac{1}{T} \Bigl\langle \textstyle\int_0^T%
    \mathrm{d}X_t \circ \nabla \phi_i(X_t)%
  \Bigr\rangle^\mathrm{neq}
\end{aligned}
\end{equation*}
with $i,\,j$ indexing the control potential basis, and the matrix $\bm{\mathsf{A}} \equiv (\mathsf{A}_{ij})$ can be inverted to solve for $\boldsymbol{\theta}^\star$ at negligible computational expense for modest basis sizes.

\bibliographystyle{apsrev}
\bibliography{main.bib}

\begin{thebibliography}{89}
\expandafter\ifx\csname natexlab\endcsname\relax\def\natexlab#1{#1}\fi
\expandafter\ifx\csname bibnamefont\endcsname\relax
  \def\bibnamefont#1{#1}\fi
\expandafter\ifx\csname bibfnamefont\endcsname\relax
  \def\bibfnamefont#1{#1}\fi
\expandafter\ifx\csname citenamefont\endcsname\relax
  \def\citenamefont#1{#1}\fi
\expandafter\ifx\csname url\endcsname\relax
  \def\url#1{\texttt{#1}}\fi
\expandafter\ifx\csname urlprefix\endcsname\relax\def\urlprefix{URL }\fi
\providecommand{\bibinfo}[2]{#2}
\providecommand{\eprint}[2][]{\url{#2}}

\bibitem[{\citenamefont{Tolman}(1938)}]{Tolman1938}
\bibinfo{author}{\bibfnamefont{R.~C.} \bibnamefont{Tolman}},
  \emph{\bibinfo{title}{{The principles of statistical mechanics}}}
  (\bibinfo{publisher}{Oxford University Press}, \bibinfo{address}{Oxford, UK},
  \bibinfo{year}{1938}), \bibinfo{edition}{1st} ed.

\bibitem[{\citenamefont{Chandler}(1987)}]{Chandler1987}
\bibinfo{author}{\bibfnamefont{D.}~\bibnamefont{Chandler}},
  \emph{\bibinfo{title}{{Introduction to modern statistical mechanics}}}
  (\bibinfo{publisher}{Oxford University Press}, \bibinfo{address}{Oxford, UK},
  \bibinfo{year}{1987}), \bibinfo{edition}{1st} ed.

\bibitem[{\citenamefont{Muegge and Hu}(2023)}]{Muegge2023}
\bibinfo{author}{\bibfnamefont{I.}~\bibnamefont{Muegge}} \bibnamefont{and}
  \bibinfo{author}{\bibfnamefont{Y.}~\bibnamefont{Hu}}, \bibinfo{journal}{ACS
  Medicinal Chemistry Letters} \textbf{\bibinfo{volume}{14}},
  \bibinfo{pages}{244} (\bibinfo{year}{2023}).

\bibitem[{\citenamefont{Cournia et~al.}(2021)\citenamefont{Cournia, Chipot,
  Roux, York, and Sherman}}]{Cournia2021}
\bibinfo{author}{\bibfnamefont{Z.}~\bibnamefont{Cournia}},
  \bibinfo{author}{\bibfnamefont{C.}~\bibnamefont{Chipot}},
  \bibinfo{author}{\bibfnamefont{B.}~\bibnamefont{Roux}},
  \bibinfo{author}{\bibfnamefont{D.~M.} \bibnamefont{York}}, \bibnamefont{and}
  \bibinfo{author}{\bibfnamefont{W.}~\bibnamefont{Sherman}}, in
  \emph{\bibinfo{booktitle}{{Free energy methods in drug discovery: Current
  state and future directions}}}, edited by
  \bibinfo{editor}{\bibfnamefont{K.~A.} \bibnamefont{Armacost}}
  \bibnamefont{and} \bibinfo{editor}{\bibfnamefont{D.~C.}
  \bibnamefont{Thompson}} (\bibinfo{publisher}{American Chemical Society},
  \bibinfo{address}{Washington, D.C.}, \bibinfo{year}{2021}),
  chap.~\bibinfo{chapter}{1}.

\bibitem[{\citenamefont{Mobley and Klimovich}(2012)}]{Mobley2012}
\bibinfo{author}{\bibfnamefont{D.~L.} \bibnamefont{Mobley}} \bibnamefont{and}
  \bibinfo{author}{\bibfnamefont{P.~V.} \bibnamefont{Klimovich}},
  \bibinfo{journal}{The Journal of Chemical Physics}
  \textbf{\bibinfo{volume}{137}}, \bibinfo{pages}{230901}
  (\bibinfo{year}{2012}).

\bibitem[{\citenamefont{Lee et~al.}(2021)\citenamefont{Lee, Kim, Cho, Lee, Lee,
  Cho, and Kim}}]{Lee2021}
\bibinfo{author}{\bibfnamefont{S.}~\bibnamefont{Lee}},
  \bibinfo{author}{\bibfnamefont{B.}~\bibnamefont{Kim}},
  \bibinfo{author}{\bibfnamefont{H.}~\bibnamefont{Cho}},
  \bibinfo{author}{\bibfnamefont{H.}~\bibnamefont{Lee}},
  \bibinfo{author}{\bibfnamefont{S.~Y.} \bibnamefont{Lee}},
  \bibinfo{author}{\bibfnamefont{E.~S.} \bibnamefont{Cho}}, \bibnamefont{and}
  \bibinfo{author}{\bibfnamefont{J.}~\bibnamefont{Kim}}, \bibinfo{journal}{ACS
  Applied Materials \& Interfaces} \textbf{\bibinfo{volume}{13}},
  \bibinfo{pages}{23647} (\bibinfo{year}{2021}).

\bibitem[{\citenamefont{Kim et~al.}(2012)\citenamefont{Kim, Lin, Martin,
  Swisher, Haranczyk, and Smit}}]{Kim2012}
\bibinfo{author}{\bibfnamefont{J.}~\bibnamefont{Kim}},
  \bibinfo{author}{\bibfnamefont{L.-C.} \bibnamefont{Lin}},
  \bibinfo{author}{\bibfnamefont{R.~L.} \bibnamefont{Martin}},
  \bibinfo{author}{\bibfnamefont{J.~A.} \bibnamefont{Swisher}},
  \bibinfo{author}{\bibfnamefont{M.}~\bibnamefont{Haranczyk}},
  \bibnamefont{and} \bibinfo{author}{\bibfnamefont{B.}~\bibnamefont{Smit}},
  \bibinfo{journal}{Langmuir} \textbf{\bibinfo{volume}{28}},
  \bibinfo{pages}{11914} (\bibinfo{year}{2012}).

\bibitem[{\citenamefont{Rickman and LeSar}(2002)}]{Rickman2002}
\bibinfo{author}{\bibfnamefont{J.~M.} \bibnamefont{Rickman}} \bibnamefont{and}
  \bibinfo{author}{\bibfnamefont{R.}~\bibnamefont{LeSar}},
  \bibinfo{journal}{Annual Review of Materials Research}
  \textbf{\bibinfo{volume}{32}}, \bibinfo{pages}{195} (\bibinfo{year}{2002}).

\bibitem[{\citenamefont{Chipot}(2023)}]{Chipot2023}
\bibinfo{author}{\bibfnamefont{C.}~\bibnamefont{Chipot}},
  \bibinfo{journal}{Annual Review of Biophysics} \textbf{\bibinfo{volume}{52}},
  \bibinfo{pages}{113} (\bibinfo{year}{2023}).

\bibitem[{\citenamefont{Axelrod et~al.}(2022)\citenamefont{Axelrod,
  Schwalbe-Koda, Mohapatra, Damewood, Greenman, and
  Gómez-Bombarelli}}]{Axelrod2022}
\bibinfo{author}{\bibfnamefont{S.}~\bibnamefont{Axelrod}},
  \bibinfo{author}{\bibfnamefont{D.}~\bibnamefont{Schwalbe-Koda}},
  \bibinfo{author}{\bibfnamefont{S.}~\bibnamefont{Mohapatra}},
  \bibinfo{author}{\bibfnamefont{J.}~\bibnamefont{Damewood}},
  \bibinfo{author}{\bibfnamefont{K.~P.} \bibnamefont{Greenman}},
  \bibnamefont{and}
  \bibinfo{author}{\bibfnamefont{R.}~\bibnamefont{Gómez-Bombarelli}},
  \bibinfo{journal}{Accounts of Materials Research}
  \textbf{\bibinfo{volume}{3}}, \bibinfo{pages}{343} (\bibinfo{year}{2022}).

\bibitem[{\citenamefont{York}(2023)}]{York2023}
\bibinfo{author}{\bibfnamefont{D.~M.} \bibnamefont{York}},
  \bibinfo{journal}{ACS Physical Chemistry Au} \textbf{\bibinfo{volume}{3}},
  \bibinfo{pages}{478} (\bibinfo{year}{2023}).

\bibitem[{\citenamefont{Dickson and Dinner}(2010)}]{Dickson2010}
\bibinfo{author}{\bibfnamefont{A.}~\bibnamefont{Dickson}} \bibnamefont{and}
  \bibinfo{author}{\bibfnamefont{A.~R.} \bibnamefont{Dinner}},
  \bibinfo{journal}{Annual Review of Physical Chemistry}
  \textbf{\bibinfo{volume}{61}}, \bibinfo{pages}{441} (\bibinfo{year}{2010}).

\bibitem[{\citenamefont{Allen et~al.}(2009)\citenamefont{Allen, Valeriani, and
  ten Wolde}}]{Allen2009}
\bibinfo{author}{\bibfnamefont{R.~J.} \bibnamefont{Allen}},
  \bibinfo{author}{\bibfnamefont{C.}~\bibnamefont{Valeriani}},
  \bibnamefont{and} \bibinfo{author}{\bibfnamefont{P.~R.} \bibnamefont{ten
  Wolde}}, \bibinfo{journal}{Journal of Physics: Condensed Matter}
  \textbf{\bibinfo{volume}{21}}, \bibinfo{pages}{463102}
  (\bibinfo{year}{2009}).

\bibitem[{\citenamefont{Zuckerman and Chong}(2017)}]{Zuckerman2017}
\bibinfo{author}{\bibfnamefont{D.~M.} \bibnamefont{Zuckerman}}
  \bibnamefont{and} \bibinfo{author}{\bibfnamefont{L.~T.} \bibnamefont{Chong}},
  \bibinfo{journal}{Annual Review of Biophysics} \textbf{\bibinfo{volume}{46}},
  \bibinfo{pages}{43} (\bibinfo{year}{2017}).

\bibitem[{\citenamefont{Hall et~al.}(2022)\citenamefont{Hall, D{\'\i}az~Leines,
  Sarupria, and Rogal}}]{hall2022practical}
\bibinfo{author}{\bibfnamefont{S.~W.} \bibnamefont{Hall}},
  \bibinfo{author}{\bibfnamefont{G.}~\bibnamefont{D{\'\i}az~Leines}},
  \bibinfo{author}{\bibfnamefont{S.}~\bibnamefont{Sarupria}}, \bibnamefont{and}
  \bibinfo{author}{\bibfnamefont{J.}~\bibnamefont{Rogal}},
  \bibinfo{journal}{The Journal of Chemical Physics}
  \textbf{\bibinfo{volume}{156}}, \bibinfo{pages}{200901}
  (\bibinfo{year}{2022}).

\bibitem[{\citenamefont{Sekimoto}(2010)}]{Sekimoto2010}
\bibinfo{author}{\bibfnamefont{K.}~\bibnamefont{Sekimoto}},
  \emph{\bibinfo{title}{{Stochastic Energetics}}} (\bibinfo{publisher}{Springer
  Berlin, Heidelberg}, \bibinfo{year}{2010}), \bibinfo{edition}{1st} ed.

\bibitem[{\citenamefont{Rold{\'a}n}(2014)}]{Roldan2014}
\bibinfo{author}{\bibfnamefont{{\'E}.}~\bibnamefont{Rold{\'a}n}},
  \emph{\bibinfo{title}{{Irreversibility and dissipation in microscopic
  systems}}} (\bibinfo{publisher}{Springer Cham}, \bibinfo{year}{2014}),
  \bibinfo{edition}{1st} ed.

\bibitem[{\citenamefont{Peliti and Pigolotti}(2021)}]{Peliti2021}
\bibinfo{author}{\bibfnamefont{L.}~\bibnamefont{Peliti}} \bibnamefont{and}
  \bibinfo{author}{\bibfnamefont{S.}~\bibnamefont{Pigolotti}},
  \emph{\bibinfo{title}{{Stochastic thermodynamics: An introduction}}}
  (\bibinfo{publisher}{Princeton University Press},
  \bibinfo{address}{Princeton, NJ}, \bibinfo{year}{2021}),
  \bibinfo{edition}{1st} ed.

\bibitem[{\citenamefont{Shiraishi}(2024)}]{Shiraishi2024}
\bibinfo{author}{\bibfnamefont{N.}~\bibnamefont{Shiraishi}},
  \emph{\bibinfo{title}{{An introduction to stochastic thermodynamics: From
  basic to advanced}}} (\bibinfo{publisher}{Springer Singapore},
  \bibinfo{year}{2024}), \bibinfo{edition}{1st} ed.

\bibitem[{\citenamefont{Limmer}(2024)}]{Limmer2024}
\bibinfo{author}{\bibfnamefont{D.~T.} \bibnamefont{Limmer}},
  \emph{\bibinfo{title}{{Statistical mechanics and stochastic thermodynamics: A
  textbook on modern approaches in and out of equilibrium}}}
  (\bibinfo{publisher}{Oxford University Press}, \bibinfo{address}{Oxford, UK},
  \bibinfo{year}{2024}), \bibinfo{edition}{1st} ed.

\bibitem[{\citenamefont{Crooks}(2001)}]{Crooks2001}
\bibinfo{author}{\bibfnamefont{G.~E.} \bibnamefont{Crooks}},
  \bibinfo{type}{{PhD thesis}}, \bibinfo{school}{University of California,
  Berkeley} (\bibinfo{year}{2001}).

\bibitem[{\citenamefont{Chetrite and Touchette}(2015)}]{Chetrite2014}
\bibinfo{author}{\bibfnamefont{R.}~\bibnamefont{Chetrite}} \bibnamefont{and}
  \bibinfo{author}{\bibfnamefont{H.}~\bibnamefont{Touchette}},
  \bibinfo{journal}{Annale Henri Poincaré} \textbf{\bibinfo{volume}{16}},
  \bibinfo{pages}{2005} (\bibinfo{year}{2015}).

\bibitem[{\citenamefont{Das and Limmer}(2019)}]{Das2019}
\bibinfo{author}{\bibfnamefont{A.}~\bibnamefont{Das}} \bibnamefont{and}
  \bibinfo{author}{\bibfnamefont{D.~T.} \bibnamefont{Limmer}},
  \bibinfo{journal}{The Journal of Chemical Physics}
  \textbf{\bibinfo{volume}{151}}, \bibinfo{pages}{244123}
  (\bibinfo{year}{2019}).

\bibitem[{\citenamefont{Das et~al.}(2022)\citenamefont{Das, Kuznets-Speck, and
  Limmer}}]{Das2022}
\bibinfo{author}{\bibfnamefont{A.}~\bibnamefont{Das}},
  \bibinfo{author}{\bibfnamefont{B.}~\bibnamefont{Kuznets-Speck}},
  \bibnamefont{and} \bibinfo{author}{\bibfnamefont{D.~T.}
  \bibnamefont{Limmer}}, \bibinfo{journal}{Physical Review Letters}
  \textbf{\bibinfo{volume}{128}}, \bibinfo{pages}{028005}
  (\bibinfo{year}{2022}).

\bibitem[{\citenamefont{Boffi and Vanden-Eijnden}(2024)}]{Boffi2024}
\bibinfo{author}{\bibfnamefont{N.~M.} \bibnamefont{Boffi}} \bibnamefont{and}
  \bibinfo{author}{\bibfnamefont{E.}~\bibnamefont{Vanden-Eijnden}},
  \bibinfo{journal}{Proceedings of the National Academy of Sciences}
  \textbf{\bibinfo{volume}{121}}, \bibinfo{pages}{e2318106121}
  (\bibinfo{year}{2024}).

\bibitem[{\citenamefont{Gibbs}(1928)}]{gibbs1928collected}
\bibinfo{author}{\bibfnamefont{J.~W.} \bibnamefont{Gibbs}},
  \emph{\bibinfo{title}{{The collected works of J. Willard Gibbs:
  Thermodynamics}}}, vol.~\bibinfo{volume}{1} (\bibinfo{publisher}{Longmans,
  Green and Company}, \bibinfo{year}{1928}).

\bibitem[{\citenamefont{Gaveau et~al.}(2002)\citenamefont{Gaveau, Martinás,
  Moreau, and Tóth}}]{Gaveau2002}
\bibinfo{author}{\bibfnamefont{B.}~\bibnamefont{Gaveau}},
  \bibinfo{author}{\bibfnamefont{K.}~\bibnamefont{Martinás}},
  \bibinfo{author}{\bibfnamefont{M.}~\bibnamefont{Moreau}}, \bibnamefont{and}
  \bibinfo{author}{\bibfnamefont{J.}~\bibnamefont{Tóth}},
  \bibinfo{journal}{Physica A: Statistical Mechanics and its Applications}
  \textbf{\bibinfo{volume}{305}}, \bibinfo{pages}{445} (\bibinfo{year}{2002}).

\bibitem[{\citenamefont{Sivak and Crooks}(2012)}]{Sivak2012}
\bibinfo{author}{\bibfnamefont{D.~A.} \bibnamefont{Sivak}} \bibnamefont{and}
  \bibinfo{author}{\bibfnamefont{G.~E.} \bibnamefont{Crooks}},
  \bibinfo{journal}{Physical Review Letters} \textbf{\bibinfo{volume}{108}},
  \bibinfo{pages}{150601} (\bibinfo{year}{2012}).

\bibitem[{\citenamefont{Kirkwood}(1935)}]{kirkwood1935statistical}
\bibinfo{author}{\bibfnamefont{J.~G.} \bibnamefont{Kirkwood}},
  \bibinfo{journal}{The Journal of Chemical Physics}
  \textbf{\bibinfo{volume}{3}}, \bibinfo{pages}{300} (\bibinfo{year}{1935}).

\bibitem[{\citenamefont{Wales and Bogdan}(2006)}]{Wales2006}
\bibinfo{author}{\bibfnamefont{D.~J.} \bibnamefont{Wales}} \bibnamefont{and}
  \bibinfo{author}{\bibfnamefont{T.~V.} \bibnamefont{Bogdan}},
  \bibinfo{journal}{The Journal of Physical Chemistry B}
  \textbf{\bibinfo{volume}{110}}, \bibinfo{pages}{20765}
  (\bibinfo{year}{2006}).

\bibitem[{\citenamefont{Baiesi and Maes}(2013)}]{baiesi2013update}
\bibinfo{author}{\bibfnamefont{M.}~\bibnamefont{Baiesi}} \bibnamefont{and}
  \bibinfo{author}{\bibfnamefont{C.}~\bibnamefont{Maes}}, \bibinfo{journal}{New
  Journal of Physics} \textbf{\bibinfo{volume}{15}}, \bibinfo{pages}{013004}
  (\bibinfo{year}{2013}).

\bibitem[{\citenamefont{Torrie and Valleau}(1977)}]{Torrie1977}
\bibinfo{author}{\bibfnamefont{G.~M.} \bibnamefont{Torrie}} \bibnamefont{and}
  \bibinfo{author}{\bibfnamefont{J.~P.} \bibnamefont{Valleau}},
  \bibinfo{journal}{Journal of Computational Physics}
  \textbf{\bibinfo{volume}{23}}, \bibinfo{pages}{187} (\bibinfo{year}{1977}).

\bibitem[{\citenamefont{Laio and Gervasio}(2008)}]{Laio2008}
\bibinfo{author}{\bibfnamefont{A.}~\bibnamefont{Laio}} \bibnamefont{and}
  \bibinfo{author}{\bibfnamefont{F.~L.} \bibnamefont{Gervasio}},
  \bibinfo{journal}{Reports on Progress in Physics}
  \textbf{\bibinfo{volume}{71}}, \bibinfo{pages}{126601}
  (\bibinfo{year}{2008}).

\bibitem[{\citenamefont{Comer et~al.}(2015)\citenamefont{Comer, Gumbart,
  Hénin, Lelièvre, Pohorille, and Chipot}}]{Comer2015}
\bibinfo{author}{\bibfnamefont{J.}~\bibnamefont{Comer}},
  \bibinfo{author}{\bibfnamefont{J.~C.} \bibnamefont{Gumbart}},
  \bibinfo{author}{\bibfnamefont{J.}~\bibnamefont{Hénin}},
  \bibinfo{author}{\bibfnamefont{T.}~\bibnamefont{Lelièvre}},
  \bibinfo{author}{\bibfnamefont{A.}~\bibnamefont{Pohorille}},
  \bibnamefont{and} \bibinfo{author}{\bibfnamefont{C.}~\bibnamefont{Chipot}},
  \bibinfo{journal}{The Journal of Physical Chemistry B}
  \textbf{\bibinfo{volume}{119}}, \bibinfo{pages}{1129} (\bibinfo{year}{2015}).

\bibitem[{\citenamefont{Lelièvre et~al.}(2010)\citenamefont{Lelièvre,
  Rousset, and Stoltz}}]{Lelievre2010}
\bibinfo{author}{\bibfnamefont{T.}~\bibnamefont{Lelièvre}},
  \bibinfo{author}{\bibfnamefont{M.}~\bibnamefont{Rousset}}, \bibnamefont{and}
  \bibinfo{author}{\bibfnamefont{G.}~\bibnamefont{Stoltz}},
  \emph{\bibinfo{title}{{Free energy computations: A mathematical
  perspective}}} (\bibinfo{publisher}{Imperial College Press},
  \bibinfo{address}{London, UK}, \bibinfo{year}{2010}), \bibinfo{edition}{1st}
  ed.

\bibitem[{\citenamefont{Frenkel and Smit}(2023)}]{frenkel2023understanding}
\bibinfo{author}{\bibfnamefont{D.}~\bibnamefont{Frenkel}} \bibnamefont{and}
  \bibinfo{author}{\bibfnamefont{B.}~\bibnamefont{Smit}},
  \emph{\bibinfo{title}{{Understanding molecular simulation: From algorithms to
  applications}}} (\bibinfo{publisher}{Academic Press, Inc.},
  \bibinfo{address}{San Diego, CA}, \bibinfo{year}{2023}),
  \bibinfo{edition}{3rd} ed.

\bibitem[{\citenamefont{Dai~Pra}(1991)}]{Pra1991}
\bibinfo{author}{\bibfnamefont{P.}~\bibnamefont{Dai~Pra}},
  \bibinfo{journal}{Applied Mathematics and Optimization}
  \textbf{\bibinfo{volume}{23}}, \bibinfo{pages}{313} (\bibinfo{year}{1991}).

\bibitem[{\citenamefont{Chetrite and Gawędzki}(2008)}]{Chetrite2008}
\bibinfo{author}{\bibfnamefont{R.}~\bibnamefont{Chetrite}} \bibnamefont{and}
  \bibinfo{author}{\bibfnamefont{K.}~\bibnamefont{Gawędzki}},
  \bibinfo{journal}{Communications in Mathematical Physics}
  \textbf{\bibinfo{volume}{282}}, \bibinfo{pages}{469} (\bibinfo{year}{2008}).

\bibitem[{\citenamefont{Gawędzki}(2013)}]{Gawedzki2013}
\bibinfo{author}{\bibfnamefont{K.}~\bibnamefont{Gawędzki}},
  \emph{\bibinfo{title}{{Fluctuation relations in stochastic thermodynamics}}}
  (\bibinfo{year}{2013}), \eprint{arXiv:1308.1518}.

\bibitem[{\citenamefont{Risken}(1996)}]{Risken1996}
\bibinfo{author}{\bibfnamefont{H.}~\bibnamefont{Risken}},
  \emph{\bibinfo{title}{{The Fokker--Planck equation: Methods of solution and
  applications}}} (\bibinfo{publisher}{Springer Berlin, Heidelberg},
  \bibinfo{year}{1996}), \bibinfo{edition}{2nd} ed.

\bibitem[{\citenamefont{Maes et~al.}(2008)\citenamefont{Maes, Neto{\v c}n{\'y},
  and Wynants}}]{Maes2008}
\bibinfo{author}{\bibfnamefont{C.}~\bibnamefont{Maes}},
  \bibinfo{author}{\bibfnamefont{K.}~\bibnamefont{Neto{\v c}n{\'y}}},
  \bibnamefont{and} \bibinfo{author}{\bibfnamefont{B.}~\bibnamefont{Wynants}},
  \bibinfo{journal}{Physica A: Statistical Mechanics and its Applications}
  \textbf{\bibinfo{volume}{387}}, \bibinfo{pages}{2675} (\bibinfo{year}{2008}).

\bibitem[{\citenamefont{Baiesi et~al.}(2009)\citenamefont{Baiesi, Maes, and
  Wynants}}]{baiesi2009fluctuations}
\bibinfo{author}{\bibfnamefont{M.}~\bibnamefont{Baiesi}},
  \bibinfo{author}{\bibfnamefont{C.}~\bibnamefont{Maes}}, \bibnamefont{and}
  \bibinfo{author}{\bibfnamefont{B.}~\bibnamefont{Wynants}},
  \bibinfo{journal}{Physical Review Letters} \textbf{\bibinfo{volume}{103}},
  \bibinfo{pages}{010602} (\bibinfo{year}{2009}).

\bibitem[{\citenamefont{Pavliotis}(2014)}]{Pavliotis2014}
\bibinfo{author}{\bibfnamefont{G.~A.} \bibnamefont{Pavliotis}},
  \emph{\bibinfo{title}{Stochastic processes and applications}}
  (\bibinfo{publisher}{Springer}, \bibinfo{address}{New York, NY},
  \bibinfo{year}{2014}), \bibinfo{edition}{1st} ed.

\bibitem[{\citenamefont{Gao and Limmer}(2019)}]{gao2019nonlinear}
\bibinfo{author}{\bibfnamefont{C.~Y.} \bibnamefont{Gao}} \bibnamefont{and}
  \bibinfo{author}{\bibfnamefont{D.~T.} \bibnamefont{Limmer}},
  \bibinfo{journal}{The Journal of Chemical Physics}
  \textbf{\bibinfo{volume}{151}}, \bibinfo{pages}{014101}
  (\bibinfo{year}{2019}).

\bibitem[{\citenamefont{Limmer et~al.}(2021)\citenamefont{Limmer, Gao, and
  Poggioli}}]{limmer2021large}
\bibinfo{author}{\bibfnamefont{D.~T.} \bibnamefont{Limmer}},
  \bibinfo{author}{\bibfnamefont{C.~Y.} \bibnamefont{Gao}}, \bibnamefont{and}
  \bibinfo{author}{\bibfnamefont{A.~R.} \bibnamefont{Poggioli}},
  \bibinfo{journal}{The European Physical Journal B}
  \textbf{\bibinfo{volume}{94}}, \bibinfo{pages}{1} (\bibinfo{year}{2021}).

\bibitem[{\citenamefont{Crooks}(1999)}]{Crooks1999}
\bibinfo{author}{\bibfnamefont{G.~E.} \bibnamefont{Crooks}},
  \bibinfo{journal}{Physical Review E} \textbf{\bibinfo{volume}{60}},
  \bibinfo{pages}{2721} (\bibinfo{year}{1999}).

\bibitem[{\citenamefont{Seifert}(2012)}]{Seifert2012}
\bibinfo{author}{\bibfnamefont{U.}~\bibnamefont{Seifert}},
  \bibinfo{journal}{Reports on Progress in Physics}
  \textbf{\bibinfo{volume}{75}}, \bibinfo{pages}{126001}
  (\bibinfo{year}{2012}).

\bibitem[{\citenamefont{Yamada and Kawasaki}(1967)}]{yamada1967nonlinear}
\bibinfo{author}{\bibfnamefont{T.}~\bibnamefont{Yamada}} \bibnamefont{and}
  \bibinfo{author}{\bibfnamefont{K.}~\bibnamefont{Kawasaki}},
  \bibinfo{journal}{Progress of Theoretical Physics}
  \textbf{\bibinfo{volume}{38}}, \bibinfo{pages}{1031} (\bibinfo{year}{1967}).

\bibitem[{\citenamefont{Morriss and Evans}(1985)}]{morriss1985isothermal}
\bibinfo{author}{\bibfnamefont{G.~P.} \bibnamefont{Morriss}} \bibnamefont{and}
  \bibinfo{author}{\bibfnamefont{D.~J.} \bibnamefont{Evans}},
  \bibinfo{journal}{Molecular Physics} \textbf{\bibinfo{volume}{54}},
  \bibinfo{pages}{629} (\bibinfo{year}{1985}).

\bibitem[{\citenamefont{Crooks}(2000)}]{crooks2000path}
\bibinfo{author}{\bibfnamefont{G.~E.} \bibnamefont{Crooks}},
  \bibinfo{journal}{Physical Review E} \textbf{\bibinfo{volume}{61}},
  \bibinfo{pages}{2361} (\bibinfo{year}{2000}).

\bibitem[{\citenamefont{Gingrich and Geissler}(2015)}]{Gingrich2015}
\bibinfo{author}{\bibfnamefont{T.~R.} \bibnamefont{Gingrich}} \bibnamefont{and}
  \bibinfo{author}{\bibfnamefont{P.~L.} \bibnamefont{Geissler}},
  \bibinfo{journal}{The Journal of Chemical Physics}
  \textbf{\bibinfo{volume}{142}}, \bibinfo{pages}{234104}
  (\bibinfo{year}{2015}).

\bibitem[{\citenamefont{Nemoto et~al.}(2016)\citenamefont{Nemoto, Bouchet,
  Jack, and Lecomte}}]{Nemoto2016}
\bibinfo{author}{\bibfnamefont{T.}~\bibnamefont{Nemoto}},
  \bibinfo{author}{\bibfnamefont{F.}~\bibnamefont{Bouchet}},
  \bibinfo{author}{\bibfnamefont{R.~L.} \bibnamefont{Jack}}, \bibnamefont{and}
  \bibinfo{author}{\bibfnamefont{V.}~\bibnamefont{Lecomte}},
  \bibinfo{journal}{Physical Review E} \textbf{\bibinfo{volume}{93}},
  \bibinfo{pages}{062123} (\bibinfo{year}{2016}).

\bibitem[{\citenamefont{Ferr\'{e} and Touchette}(2018)}]{Ferre2018}
\bibinfo{author}{\bibfnamefont{G.}~\bibnamefont{Ferr\'{e}}} \bibnamefont{and}
  \bibinfo{author}{\bibfnamefont{H.}~\bibnamefont{Touchette}},
  \bibinfo{journal}{Journal of Statistical Physics}
  \textbf{\bibinfo{volume}{172}}, \bibinfo{pages}{1525} (\bibinfo{year}{2018}).

\bibitem[{\citenamefont{Ferr\'{e} and Grafke}(2021)}]{Ferre2021}
\bibinfo{author}{\bibfnamefont{G.}~\bibnamefont{Ferr\'{e}}} \bibnamefont{and}
  \bibinfo{author}{\bibfnamefont{T.}~\bibnamefont{Grafke}},
  \bibinfo{journal}{Multiscale Modeling \& Simulation}
  \textbf{\bibinfo{volume}{19}}, \bibinfo{pages}{1310} (\bibinfo{year}{2021}).

\bibitem[{\citenamefont{Kuznets-Speck and
  Limmer}(2021)}]{kuznets2021dissipation}
\bibinfo{author}{\bibfnamefont{B.}~\bibnamefont{Kuznets-Speck}}
  \bibnamefont{and} \bibinfo{author}{\bibfnamefont{D.~T.}
  \bibnamefont{Limmer}}, \bibinfo{journal}{Proceedings of the National Academy
  of Sciences} \textbf{\bibinfo{volume}{118}}, \bibinfo{pages}{e2020863118}
  (\bibinfo{year}{2021}).

\bibitem[{\citenamefont{Kuznets-Speck and Limmer}(2023)}]{kuznets2023inferring}
\bibinfo{author}{\bibfnamefont{B.}~\bibnamefont{Kuznets-Speck}}
  \bibnamefont{and} \bibinfo{author}{\bibfnamefont{D.~T.}
  \bibnamefont{Limmer}}, \bibinfo{journal}{Biophysical Journal}
  \textbf{\bibinfo{volume}{122}}, \bibinfo{pages}{1659} (\bibinfo{year}{2023}).

\bibitem[{\citenamefont{Lesnicki et~al.}(2020)\citenamefont{Lesnicki, Gao,
  Rotenberg, and Limmer}}]{lesnicki2020field}
\bibinfo{author}{\bibfnamefont{D.}~\bibnamefont{Lesnicki}},
  \bibinfo{author}{\bibfnamefont{C.~Y.} \bibnamefont{Gao}},
  \bibinfo{author}{\bibfnamefont{B.}~\bibnamefont{Rotenberg}},
  \bibnamefont{and} \bibinfo{author}{\bibfnamefont{D.~T.}
  \bibnamefont{Limmer}}, \bibinfo{journal}{Physical Review Letters}
  \textbf{\bibinfo{volume}{124}}, \bibinfo{pages}{206001}
  (\bibinfo{year}{2020}).

\bibitem[{\citenamefont{Lesnicki et~al.}(2021)\citenamefont{Lesnicki, Gao,
  Limmer, and Rotenberg}}]{lesnicki2021molecular}
\bibinfo{author}{\bibfnamefont{D.}~\bibnamefont{Lesnicki}},
  \bibinfo{author}{\bibfnamefont{C.~Y.} \bibnamefont{Gao}},
  \bibinfo{author}{\bibfnamefont{D.~T.} \bibnamefont{Limmer}},
  \bibnamefont{and}
  \bibinfo{author}{\bibfnamefont{B.}~\bibnamefont{Rotenberg}},
  \bibinfo{journal}{The Journal of Chemical Physics}
  \textbf{\bibinfo{volume}{155}}, \bibinfo{pages}{014507}
  (\bibinfo{year}{2021}).

\bibitem[{\citenamefont{Hummer}(2007)}]{Hummer2007}
\bibinfo{author}{\bibfnamefont{G.}~\bibnamefont{Hummer}}, in
  \emph{\bibinfo{booktitle}{{Free energy calculations: Theory and applications
  in chemistry and biology}}}, edited by
  \bibinfo{editor}{\bibfnamefont{C.}~\bibnamefont{Chipot}} \bibnamefont{and}
  \bibinfo{editor}{\bibfnamefont{A.}~\bibnamefont{Porohille}}
  (\bibinfo{publisher}{Springer Berlin, Heidelberg}, \bibinfo{year}{2007}),
  chap.~\bibinfo{chapter}{5}.

\bibitem[{\citenamefont{Øksendal}(2010)}]{Oksendal2003}
\bibinfo{author}{\bibfnamefont{B.}~\bibnamefont{Øksendal}},
  \emph{\bibinfo{title}{{Stochastic differential equations: An introduction
  with applications}}} (\bibinfo{publisher}{Springer Berlin, Heidelberg},
  \bibinfo{year}{2010}), \bibinfo{edition}{6th} ed.

\bibitem[{\citenamefont{Bierkens and Kappen}(2014)}]{Bierkens2014}
\bibinfo{author}{\bibfnamefont{J.}~\bibnamefont{Bierkens}} \bibnamefont{and}
  \bibinfo{author}{\bibfnamefont{H.~J.} \bibnamefont{Kappen}},
  \bibinfo{journal}{Systems \& Control Letters} \textbf{\bibinfo{volume}{72}},
  \bibinfo{pages}{36} (\bibinfo{year}{2014}).

\bibitem[{\citenamefont{Thijssen and Kappen}(2015)}]{Thijssen2015}
\bibinfo{author}{\bibfnamefont{S.}~\bibnamefont{Thijssen}} \bibnamefont{and}
  \bibinfo{author}{\bibfnamefont{H.~J.} \bibnamefont{Kappen}},
  \bibinfo{journal}{Physical Review E} \textbf{\bibinfo{volume}{91}},
  \bibinfo{pages}{032104} (\bibinfo{year}{2015}).

\bibitem[{\citenamefont{Sutton and Barto}(2018)}]{Sutton2018}
\bibinfo{author}{\bibfnamefont{R.~S.} \bibnamefont{Sutton}} \bibnamefont{and}
  \bibinfo{author}{\bibfnamefont{A.~G.} \bibnamefont{Barto}},
  \emph{\bibinfo{title}{{Reinforcement learning: An introduction}}}
  (\bibinfo{publisher}{The MIT Press}, \bibinfo{address}{Cambridge, MA},
  \bibinfo{year}{2018}), \bibinfo{edition}{2nd} ed.

\bibitem[{\citenamefont{Rose et~al.}(2021)\citenamefont{Rose, Mair, and
  Garrahan}}]{rose2021reinforcement}
\bibinfo{author}{\bibfnamefont{D.~C.} \bibnamefont{Rose}},
  \bibinfo{author}{\bibfnamefont{J.~F.} \bibnamefont{Mair}}, \bibnamefont{and}
  \bibinfo{author}{\bibfnamefont{J.~P.} \bibnamefont{Garrahan}},
  \bibinfo{journal}{New Journal of Physics} \textbf{\bibinfo{volume}{23}},
  \bibinfo{pages}{013013} (\bibinfo{year}{2021}).

\bibitem[{\citenamefont{Das et~al.}(2021)\citenamefont{Das, Rose, Garrahan, and
  Limmer}}]{Das2021b}
\bibinfo{author}{\bibfnamefont{A.}~\bibnamefont{Das}},
  \bibinfo{author}{\bibfnamefont{D.~C.} \bibnamefont{Rose}},
  \bibinfo{author}{\bibfnamefont{J.~P.} \bibnamefont{Garrahan}},
  \bibnamefont{and} \bibinfo{author}{\bibfnamefont{D.~T.}
  \bibnamefont{Limmer}}, \bibinfo{journal}{The Journal of Chemical Physics}
  \textbf{\bibinfo{volume}{155}}, \bibinfo{pages}{134105}
  (\bibinfo{year}{2021}).

\bibitem[{\citenamefont{Yan et~al.}(2022)\citenamefont{Yan, Touchette, and
  Rotskoff}}]{yan2022learning}
\bibinfo{author}{\bibfnamefont{J.}~\bibnamefont{Yan}},
  \bibinfo{author}{\bibfnamefont{H.}~\bibnamefont{Touchette}},
  \bibnamefont{and} \bibinfo{author}{\bibfnamefont{G.~M.}
  \bibnamefont{Rotskoff}}, \bibinfo{journal}{Physical Review E}
  \textbf{\bibinfo{volume}{105}}, \bibinfo{pages}{024115}
  (\bibinfo{year}{2022}).

\bibitem[{\citenamefont{Warren and Allen}(2014)}]{Warren2014}
\bibinfo{author}{\bibfnamefont{P.~B.} \bibnamefont{Warren}} \bibnamefont{and}
  \bibinfo{author}{\bibfnamefont{R.~J.} \bibnamefont{Allen}},
  \bibinfo{journal}{Entropy} \textbf{\bibinfo{volume}{16}},
  \bibinfo{pages}{221} (\bibinfo{year}{2014}).

\bibitem[{\citenamefont{Singh and Limmer}(2023)}]{singh2023variational}
\bibinfo{author}{\bibfnamefont{A.~N.} \bibnamefont{Singh}} \bibnamefont{and}
  \bibinfo{author}{\bibfnamefont{D.~T.} \bibnamefont{Limmer}},
  \bibinfo{journal}{The Journal of Chemical Physics}
  \textbf{\bibinfo{volume}{159}}, \bibinfo{pages}{024124}
  (\bibinfo{year}{2023}).

\bibitem[{\citenamefont{Jung et~al.}(2023)\citenamefont{Jung, Covino, Arjun,
  Leitold, Dellago, Bolhuis, and Hummer}}]{jung2023machine}
\bibinfo{author}{\bibfnamefont{H.}~\bibnamefont{Jung}},
  \bibinfo{author}{\bibfnamefont{R.}~\bibnamefont{Covino}},
  \bibinfo{author}{\bibfnamefont{A.}~\bibnamefont{Arjun}},
  \bibinfo{author}{\bibfnamefont{C.}~\bibnamefont{Leitold}},
  \bibinfo{author}{\bibfnamefont{C.}~\bibnamefont{Dellago}},
  \bibinfo{author}{\bibfnamefont{P.~G.} \bibnamefont{Bolhuis}},
  \bibnamefont{and} \bibinfo{author}{\bibfnamefont{G.}~\bibnamefont{Hummer}},
  \bibinfo{journal}{Nature Computational Science} \textbf{\bibinfo{volume}{3}},
  \bibinfo{pages}{334} (\bibinfo{year}{2023}).

\bibitem[{\citenamefont{Durrett}(2010)}]{Durrett2010}
\bibinfo{author}{\bibfnamefont{R.}~\bibnamefont{Durrett}},
  \emph{\bibinfo{title}{{Probability: Theory and Examples}}}
  (\bibinfo{publisher}{Cambridge University Press},
  \bibinfo{address}{Cambridge, UK}, \bibinfo{year}{2010}),
  \bibinfo{edition}{4th} ed.

\bibitem[{\citenamefont{Boffi and Vanden-Eijnden}(2023)}]{Boffi2023}
\bibinfo{author}{\bibfnamefont{N.~M.} \bibnamefont{Boffi}} \bibnamefont{and}
  \bibinfo{author}{\bibfnamefont{E.}~\bibnamefont{Vanden-Eijnden}},
  \bibinfo{journal}{Machine Learning: Science and Technology}
  \textbf{\bibinfo{volume}{4}}, \bibinfo{pages}{035012} (\bibinfo{year}{2023}).

\bibitem[{\citenamefont{Kloeden and Platen}(1992)}]{Kloeden1992}
\bibinfo{author}{\bibfnamefont{P.~E.} \bibnamefont{Kloeden}} \bibnamefont{and}
  \bibinfo{author}{\bibfnamefont{E.}~\bibnamefont{Platen}},
  \emph{\bibinfo{title}{{Numerical solution of stochastic differential
  equations}}} (\bibinfo{publisher}{Springer Berlin, Heidelberg},
  \bibinfo{year}{1992}), \bibinfo{edition}{1st} ed.

\bibitem[{\citenamefont{Doi and Edwards}(1988)}]{doi1988theory}
\bibinfo{author}{\bibfnamefont{M.}~\bibnamefont{Doi}} \bibnamefont{and}
  \bibinfo{author}{\bibfnamefont{S.~F.} \bibnamefont{Edwards}},
  \emph{\bibinfo{title}{{The theory of polymer dynamics}}}
  (\bibinfo{publisher}{Oxford University Press}, \bibinfo{address}{Oxford, UK},
  \bibinfo{year}{1988}), \bibinfo{edition}{2nd} ed.

\bibitem[{\citenamefont{Weeks et~al.}(1971)\citenamefont{Weeks, Chandler, and
  Andersen}}]{weeks1971role}
\bibinfo{author}{\bibfnamefont{J.~D.} \bibnamefont{Weeks}},
  \bibinfo{author}{\bibfnamefont{D.}~\bibnamefont{Chandler}}, \bibnamefont{and}
  \bibinfo{author}{\bibfnamefont{H.~C.} \bibnamefont{Andersen}},
  \bibinfo{journal}{The Journal of Chemical Physics}
  \textbf{\bibinfo{volume}{54}}, \bibinfo{pages}{5237} (\bibinfo{year}{1971}).

\bibitem[{\citenamefont{Bianco et~al.}(2018)\citenamefont{Bianco, Locatelli,
  and Malgaretti}}]{Bianco2018}
\bibinfo{author}{\bibfnamefont{V.}~\bibnamefont{Bianco}},
  \bibinfo{author}{\bibfnamefont{E.}~\bibnamefont{Locatelli}},
  \bibnamefont{and}
  \bibinfo{author}{\bibfnamefont{P.}~\bibnamefont{Malgaretti}},
  \bibinfo{journal}{Physical Review Letters} \textbf{\bibinfo{volume}{121}},
  \bibinfo{pages}{217802} (\bibinfo{year}{2018}).

\bibitem[{\citenamefont{Anand and Singh}(2018)}]{Anand2018}
\bibinfo{author}{\bibfnamefont{S.~K.} \bibnamefont{Anand}} \bibnamefont{and}
  \bibinfo{author}{\bibfnamefont{S.~P.} \bibnamefont{Singh}},
  \bibinfo{journal}{Physical Review E} \textbf{\bibinfo{volume}{98}},
  \bibinfo{pages}{042501} (\bibinfo{year}{2018}).

\bibitem[{\citenamefont{Kumar et~al.}(1992)\citenamefont{Kumar, Rosenberg,
  Bouzida, Swendsen, and Kollman}}]{Kumar1992}
\bibinfo{author}{\bibfnamefont{S.}~\bibnamefont{Kumar}},
  \bibinfo{author}{\bibfnamefont{J.~M.} \bibnamefont{Rosenberg}},
  \bibinfo{author}{\bibfnamefont{D.}~\bibnamefont{Bouzida}},
  \bibinfo{author}{\bibfnamefont{R.~H.} \bibnamefont{Swendsen}},
  \bibnamefont{and} \bibinfo{author}{\bibfnamefont{P.~A.}
  \bibnamefont{Kollman}}, \bibinfo{journal}{Journal of Computational Chemistry}
  \textbf{\bibinfo{volume}{13}}, \bibinfo{pages}{1011} (\bibinfo{year}{1992}).

\bibitem[{\citenamefont{Roux}(1995)}]{Roux1995}
\bibinfo{author}{\bibfnamefont{B.}~\bibnamefont{Roux}},
  \bibinfo{journal}{Computer Physics Communications}
  \textbf{\bibinfo{volume}{91}}, \bibinfo{pages}{275} (\bibinfo{year}{1995}).

\bibitem[{\citenamefont{Anderson}(1982)}]{Anderson1982}
\bibinfo{author}{\bibfnamefont{{\relax B.\,D.\,O}.}~\bibnamefont{Anderson}},
  \bibinfo{journal}{Stochastic Processes and their Applications}
  \textbf{\bibinfo{volume}{12}}, \bibinfo{pages}{313} (\bibinfo{year}{1982}).

\bibitem[{\citenamefont{Pavon}(1989)}]{Pavon1989}
\bibinfo{author}{\bibfnamefont{M.}~\bibnamefont{Pavon}},
  \bibinfo{journal}{Applied Mathematics and Optimization}
  \textbf{\bibinfo{volume}{19}}, \bibinfo{pages}{187} (\bibinfo{year}{1989}).

\bibitem[{\citenamefont{Chetrite and Gupta}(2011)}]{Chetrite2011}
\bibinfo{author}{\bibfnamefont{R.}~\bibnamefont{Chetrite}} \bibnamefont{and}
  \bibinfo{author}{\bibfnamefont{S.}~\bibnamefont{Gupta}},
  \bibinfo{journal}{Journal of Statistical Physics}
  \textbf{\bibinfo{volume}{143}}, \bibinfo{pages}{543} (\bibinfo{year}{2011}).

\bibitem[{\citenamefont{Sohl-Dickstein
  et~al.}(2015)\citenamefont{Sohl-Dickstein, Weiss, Maheswaranathan, and
  Ganguli}}]{Sohl-Dickstein2015}
\bibinfo{author}{\bibfnamefont{J.}~\bibnamefont{Sohl-Dickstein}},
  \bibinfo{author}{\bibfnamefont{E.}~\bibnamefont{Weiss}},
  \bibinfo{author}{\bibfnamefont{N.}~\bibnamefont{Maheswaranathan}},
  \bibnamefont{and} \bibinfo{author}{\bibfnamefont{S.}~\bibnamefont{Ganguli}},
  in \emph{\bibinfo{booktitle}{Proceedings of the 32nd International Conference
  on Machine Learning}}, edited by
  \bibinfo{editor}{\bibfnamefont{F.}~\bibnamefont{Bach}} \bibnamefont{and}
  \bibinfo{editor}{\bibfnamefont{D.}~\bibnamefont{Blei}}
  (\bibinfo{publisher}{PMLR}, \bibinfo{address}{Lille, France},
  \bibinfo{year}{2015}), vol.~\bibinfo{volume}{37} of
  \emph{\bibinfo{series}{Proceedings of Machine Learning Research}}, p.
  \bibinfo{pages}{2256}.

\bibitem[{\citenamefont{Song et~al.}(2021)\citenamefont{Song, Sohl{-}Dickstein,
  Kingma, Kumar, Ermon, and Poole}}]{Song2021amended}
\bibinfo{author}{\bibfnamefont{Y.}~\bibnamefont{Song}},
  \bibinfo{author}{\bibfnamefont{J.}~\bibnamefont{Sohl{-}Dickstein}},
  \bibinfo{author}{\bibfnamefont{D.~P.} \bibnamefont{Kingma}},
  \bibinfo{author}{\bibfnamefont{A.}~\bibnamefont{Kumar}},
  \bibinfo{author}{\bibfnamefont{S.}~\bibnamefont{Ermon}}, \bibnamefont{and}
  \bibinfo{author}{\bibfnamefont{B.}~\bibnamefont{Poole}}, in
  \emph{\bibinfo{booktitle}{International Conference on Learning
  Representations}} (\bibinfo{year}{2021}).

\bibitem[{\citenamefont{De~Bortoli et~al.}(2021)\citenamefont{De~Bortoli,
  Thornton, Heng, and Doucet}}]{DeBortoli2021}
\bibinfo{author}{\bibfnamefont{V.}~\bibnamefont{De~Bortoli}},
  \bibinfo{author}{\bibfnamefont{J.}~\bibnamefont{Thornton}},
  \bibinfo{author}{\bibfnamefont{J.}~\bibnamefont{Heng}}, \bibnamefont{and}
  \bibinfo{author}{\bibfnamefont{A.}~\bibnamefont{Doucet}}, in
  \emph{\bibinfo{booktitle}{Advances in Neural Information Processing
  Systems}}, edited by
  \bibinfo{editor}{\bibfnamefont{M.}~\bibnamefont{Ranzato}},
  \bibinfo{editor}{\bibfnamefont{A.}~\bibnamefont{Beygelzimer}},
  \bibinfo{editor}{\bibfnamefont{Y.}~\bibnamefont{Dauphin}},
  \bibinfo{editor}{\bibfnamefont{P.}~\bibnamefont{Liang}}, \bibnamefont{and}
  \bibinfo{editor}{\bibfnamefont{J.~W.} \bibnamefont{Vaughan}}
  (\bibinfo{publisher}{Curran Associates, Inc.}, \bibinfo{address}{Red Hook,
  NY}, \bibinfo{year}{2021}), vol.~\bibinfo{volume}{34}, p.
  \bibinfo{pages}{17695}.

\bibitem[{\citenamefont{Prato}(2014)}]{Prato2014}
\bibinfo{author}{\bibfnamefont{G.}~\bibnamefont{Prato}},
  \emph{\bibinfo{title}{Introduction to stochastic analysis and Malliavin
  calculus}} (\bibinfo{publisher}{Edizione della Normale Pisa},
  \bibinfo{year}{2014}), \bibinfo{edition}{1st} ed.

\bibitem[{\citenamefont{Ray et~al.}(2018{\natexlab{a}})\citenamefont{Ray, Chan,
  and Limmer}}]{Ray2018a}
\bibinfo{author}{\bibfnamefont{U.}~\bibnamefont{Ray}},
  \bibinfo{author}{\bibfnamefont{G.~K.-L.} \bibnamefont{Chan}},
  \bibnamefont{and} \bibinfo{author}{\bibfnamefont{D.~T.}
  \bibnamefont{Limmer}}, \bibinfo{journal}{The Journal of Chemical Physics}
  \textbf{\bibinfo{volume}{148}}, \bibinfo{pages}{124120}
  (\bibinfo{year}{2018}{\natexlab{a}}).

\bibitem[{\citenamefont{Ray et~al.}(2018{\natexlab{b}})\citenamefont{Ray, Chan,
  and Limmer}}]{Ray2018b}
\bibinfo{author}{\bibfnamefont{U.}~\bibnamefont{Ray}},
  \bibinfo{author}{\bibfnamefont{G.~K.-L.} \bibnamefont{Chan}},
  \bibnamefont{and} \bibinfo{author}{\bibfnamefont{D.~T.}
  \bibnamefont{Limmer}}, \bibinfo{journal}{Physical Review Letters}
  \textbf{\bibinfo{volume}{120}}, \bibinfo{pages}{210602}
  (\bibinfo{year}{2018}{\natexlab{b}}).

\bibitem[{\citenamefont{Ray and Chan}(2020)}]{Ray2020}
\bibinfo{author}{\bibfnamefont{U.}~\bibnamefont{Ray}} \bibnamefont{and}
  \bibinfo{author}{\bibfnamefont{G.~K.-L.} \bibnamefont{Chan}},
  \bibinfo{journal}{The Journal of Chemical Physics}
  \textbf{\bibinfo{volume}{152}}, \bibinfo{pages}{104107}
  (\bibinfo{year}{2020}).

\bibitem[{\citenamefont{Angeli et~al.}(2021)\citenamefont{Angeli, Grosskinsky,
  and Johansen}}]{Angeli2021}
\bibinfo{author}{\bibfnamefont{L.}~\bibnamefont{Angeli}},
  \bibinfo{author}{\bibfnamefont{S.}~\bibnamefont{Grosskinsky}},
  \bibnamefont{and} \bibinfo{author}{\bibfnamefont{A.~M.}
  \bibnamefont{Johansen}}, \bibinfo{journal}{Stochastic Processes and their
  Applications} \textbf{\bibinfo{volume}{138}}, \bibinfo{pages}{117}
  (\bibinfo{year}{2021}).

\end{thebibliography}

\end{document}